\documentclass[english,aps,prb,notitlepage,twocolumn,%
nofootinbib,
longbibliography,superscriptaddress,secnumarabic,floatfix]{revtex4}

\usepackage[final]{graphicx}%

\usepackage{color}
\usepackage[usenames,dvipsnames]{xcolor}
\usepackage{hyperref} 
\hypersetup{%
	linktoc=page,%
	colorlinks=true,%
	linkcolor=NavyBlue,%
	citecolor=NavyBlue,%
	urlcolor=NavyBlue,%
	linkbordercolor=red,%
	pdfborderstyle={/S/U/W 1},%
}
\usepackage{amssymb,amsmath,bm,soul}
\setul{1.4pt}{.4pt}
\newcommand{\uhref}[2]{{\href{#1}{\ul{#2}}}}

\renewcommand{\vec}{\boldsymbol}
\newcommand{\C}{\bar{C}}
\newcommand{\sss}{\scriptscriptstyle}

\newcommand{\eq}{Eq.\ }
\newcommand{\eqs}{Eqs.\ }

\newcommand{\sbonlinecite}[1]{[\onlinecite{#1}]}
\newcommand{\notll}{\hskip 0.4mm \not \hskip -0.4mm \ll}

\newcommand{\eone}{\varepsilon_{1}}

\begin{document}
\title{Peak effect due to competing vortex ground states in superconductors with large inclusions}
	
\author{Roland Willa}
\affiliation{
	Materials Science Division,
	Argonne National Laboratory,
	9700 South Cass Av, Argonne, IL 60639, USA
}

\author{Alexei E. Koshelev}
\affiliation{
	Materials Science Division,
	Argonne National Laboratory,
	9700 South Cass Av, Argonne, IL 60639, USA
}
	
\author{Ivan A. Sadovskyy}
\affiliation{
	Materials Science Division,
	Argonne National Laboratory,
	9700 South Cass Av, Argonne, IL 60639, USA
}
\affiliation{
    Computation Institute,
    University of Chicago,
    5735 South Ellis Av, Chicago, IL 60637, USA
}

\author{Andreas Glatz}
\affiliation{
	Materials Science Division,
	Argonne National Laboratory,
	9700 South Cass Av, Argonne, IL 60639, USA
}
\affiliation{
	Department of Physics,
	Northern Illinois University,
	DeKalb, IL 60115, USA
}

\date{\today}

\begin{abstract}	
Superconductors can support large dissipation-free electrical currents only if vortex lines are effectively immobilized by material defects. Macroscopic critical currents depend on elemental interactions of vortices with individual pinning centers. Pinning mechanisms are nontrivial for large-size defects such as self-assembled nanoparticles. We investigate the problem of a vortex system  interacting with an isolated defect using time-dependent Ginzburg-Landau simulations. In particular, we study the instability-limited depinning process and extract the dependence of the pin-breaking force on inclusion size and anisotropy for an \emph{isolated vortex line}. In the case of a \emph{vortex lattice} interacting with a large isolated defect, we find a series of first-order phase transitions at well-defined magnetic fields, when the number of vortex lines occupying the inclusion changes. The pin-breaking force has sharp local minima at those fields. As a consequence, in the case of isolated identical large-size defects, the field dependence of the critical current is composed of a series of peaks located in between the occupation-number transition points.
\end{abstract}

\maketitle

\section{Introduction}

The electrodynamic properties of type-II superconductors are mostly determined by vortex lines\cite{Abrikosov1957}, tubes carrying a quantized magnetic flux $\Phi_{0} = hc/2e$ screened by circulating supercurrents. Effective immobilization of vortices by material defects is essential for the ability of practical superconductors to carry large electrical currents without dissipation.  

Recently, impressive progress has been made in the controlled fabrication of materials containing defect structures which provide effective pinning landscapes, see reviews~[\onlinecite{Matsumoto2009, Obradors2014, GurevichAnnRevCMP14, KwokRoPP2016, FeighanSUST17}]. The most prominent  example  is the synthesis of cuprate high-temperature superconductors containing self-assembled oxide  precipitates in the form of nanoparticles\cite{MacManusAPL04, HauganNat04, SongAPL06, GutierrezNatMat07, YamasakiSUST08, YamasakiSUST16, PolatPhysRevB11, MiuraPhysRevB11, Miura2013, MiuraNPGAsMat17} or nanorods.\cite{GoyalSUST05, KangSci06, Mele2008, Selva2015b, AwajiSuST2017} In this paper we focus on  materials in which the dominant pinning centers are large-size nanoparticles. Another promising technique to generate particle-like pinning centers in the form of small clusters, is proton or ion irradiation.\cite{JiaAPL13, Kihlstrom2013, Haberkorn2015, Leroux2015, Taen2012, Taen2015} Despite these practical advances, the understanding of pinning mechanisms in those materials remains unsatisfactory.

The development of qualitative and quantitative descriptions of vortex pinning by point defects has been subject to intense theoretical\cite{Labusch1969, LarkinO:1979, Vinokur1990, Coffey1991, OvchinnikovI:1991, BlatterFGLV:1994, Brandt:1995, BlatterGK:2004, GurevichSST07, Buchacek2018a} and numerical \cite{BrandtJLTP83-1, BrandtJLTP83-2, JensenPhysRevLett88, KoshelevPhysRevLett94, OtterloPRL00, WinieckiA:2002, LuoHu:2007, LuoHuJSNM10, Koshelev:2011, DobramyslEPJ13, KoshelevPRB16, Willa2018a}  research over the past decades. Theoretical studies are mostly focused on two major topics: weak collective pinning by large number of atomic defects\cite{LarkinO:1979, BlatterFGLV:1994, Brandt:1995} or strong pinning by dilute distribution of defects.\cite{Labusch1969, LarkinO:1979, OvchinnikovI:1991, BlatterGK:2004, Buchacek2018a} The second scenario is most relevant for  superconducting materials in which nanoparticles are the main pinning centers. Hereby, the elemental interaction of a vortex line with a single defect constitutes the basic ingredient for the theoretical description. This microscopic interaction is characterized by two pinning parameters: the pin-breaking force and the associated pinning energy. In a conventional approach, where one assumes strong but \emph{point-like} defects, both quantities can be evaluated quantitatively.
\begin{figure}[b]
\centering	
\includegraphics[width = .48\textwidth]{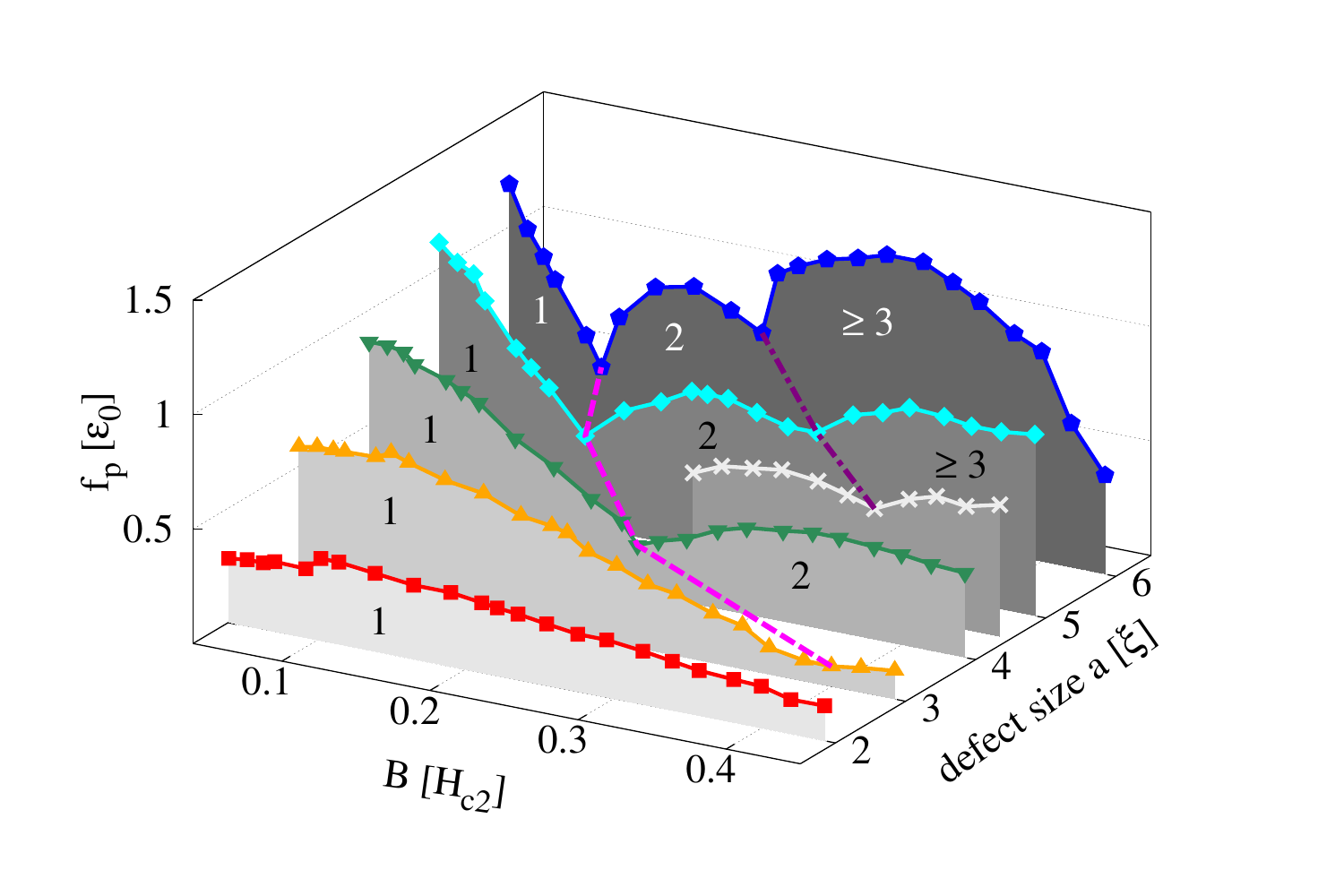}
\caption{Critical force $f_{p}$ to depin a vortex lattice from an isolated defect as a function of field strength $B$ and  defect diameter $a$. At a specific field $B_2(a)$ (dashed, magenta line) double-occupation of the defect becomes favorable as compared to single occupation. At this transition, $f_{p}(B)$ shows a sharp cusp. For large defects a second cusp at $B_{3}(a)$ (dot-dashed, purple line) marks the transition to triple-occupancy of the defect.
}
\label{fig:fp-of-a-B}
\end{figure}

Recently, large-scale simulations of the time-dependent Ginzburg-Landau model have been demonstrated to be very useful for exploring different pinning regimes. \cite{SadovskyyJComp2015, KoshelevPRB16, SadovskyyPRAppl2016, SadovskyyAdvMat2016, KimmelPRE2017, Willa2018a} In this paper we use this approach to investigate a single vortex line (zero-field limit) and a vortex lattice at finite magnetic field interacting with an \emph{isolated} large-size inclusion. In contrast to small defects, the depinning from a large defect is a nontrivial process strongly influenced by elastic deformation of the vortex line outside defect. We quantify this process and compute the dependence of the pin-breaking force on the inclusion size and on the material's anisotropy. 

It is generally believed that mostly \emph{intrinsic} defect properties (its size, shape, internal structure) determine the pinning parameters, and hence, they are usually assumed to be field independent \cite{LarkinO:1979, BlatterFGLV:1994, BlatterGK:2004}. We find, however, that this assumption is strongly violated for large inclusions. By studying the simple setting of an isolated defect interacting with a vortex lattice, we show that surrounding vortex lines substantially influence the pinning/depinning process leading to the strong dependence of the pin-breaking force on the field strength. This means that, in general, the pin-breaking force is \emph{not an intrinsic} defect property. Moreover, at high magnetic fields, the inclusion may capture more than one vortex. Each increase of the inclusion's occupation number corresponds to a first-order phase transition between pinning ground states. This manifests itself through pronounced features in the pin-breaking force. The field dependences of this key parameter for inclusions with different sizes, shown in Fig.~\ref{fig:fp-of-a-B}, highlight the main phenomenon discussed here: we report a peak effect associated with transitions of the defect's pinning ground-states between different (vortex) occupancies.

The paper is organized as follows: In section~\ref{Sec:singleIncl} the critical state of a single flux line detaching from an inclusion is investigated as function of the defect's size and of the superconductor's uniaxial anisotropy. In section~\ref{sec:manyvortices-singledefect} we review existing theoretical descriptions of the single-defect problem, discuss possible limitations, and propose generalizations. We then study the capability of a single defect to pin a vortex lattice by numerical means. In particular, the field strength $B$ [or equivalently the intervortex distance $a_{\sss \triangle} = (4/3)^{1/4} (\Phi_{0}/B)^{1/2}$] is shown to play an important role. While details about the implementation of the numerical solver for the Ginzburg-Landau equation on graphics processing units (GPUs) is published elsewhere,\cite{SadovskyyJComp2015} the specifications used in this work are discussed in Appendix~\ref{Sec:model}. Appendix~\ref{sec:lin-el} is devoted to the vortex lattice's response to small external forces (elastic regime). While the main text primarily focuses on the critical current, a discussion of another observable---the $ac$ penetration depth (or Campbell length) of a low-frequency field oscillation---shall be given in Appendix~\ref{sec:Campbell}.

\section{Interaction of a single vortex with an isolated spherical inclusion}
\label{Sec:singleIncl}
\begin{figure}[tb]
\centering
\includegraphics[width= .48\textwidth]{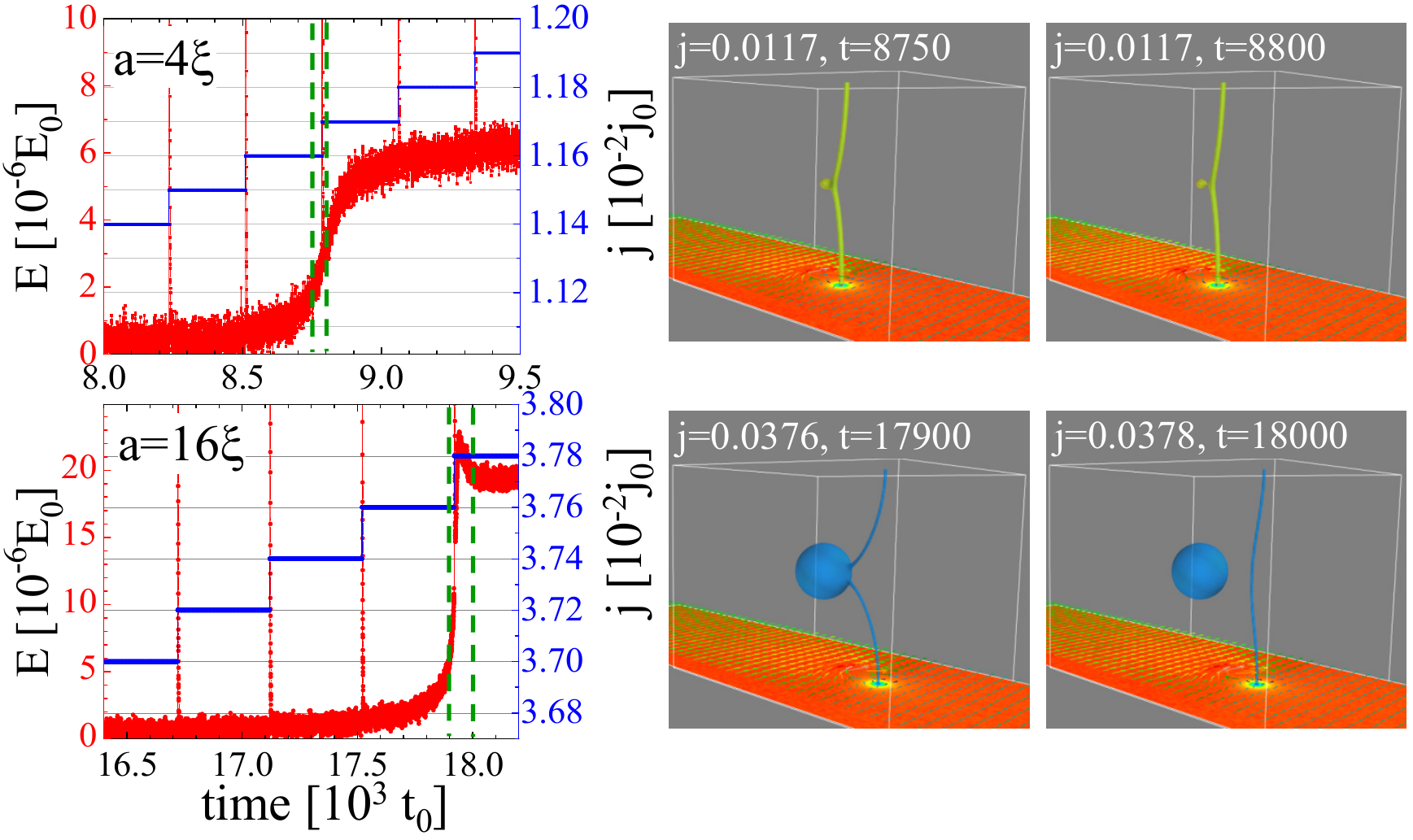}
\caption{Time dependence of the applied current (blue steps) and the resulting electric field (red) from TDGL simulations of a single vortex line pinned at an isolated inclusion. The data is shown for spherical inclusions with diameters $4 \xi$ and $16 \xi$. At specific times [marked by vertical dashed lines (green)], snapshots of the vortex configurations are shown. The units of time $t_0$, electric field $E_0$, and current $j_0$ are defined in the Appendix \ref{Sec:model}.
}
\label{Fig-ParticlePinBreak}	
\end{figure}

In this section we investigate the size- and anisotropy-dependences of the maximum (or pin-breaking) force for an isolated vortex line depinning from a spherical defect. To evaluate this key parameter characterizing an individual pinning center, we will numerically compute the critical current $j_{c}$ necessary to depin a single vortex from an isolated defect. The current $j$ (applied along $y$) acts on the flux line (threading along $z$) with the total Lorentz force $f_{L} = \Phi_{0} j L_{z} /c$ (along $x$) over the vortex' entire length $L_{z}$ (here $c$ is the speed of light; we use CGS units). For small currents%
\footnote{Note, that the attribute \emph{small} used for the current strength depends on the system's size. For a single defect the critical current scales as $j_{c} = f_{p} c / (\Phi_{0} L_{z})$, while for finite fields it reads $j_{c} = f_{p} c / (B L_{x}L_{y}L_{z})$.}%
, $j<j_{c}$, the vortex remains trapped in the inclusion by a counteracting pinning force but deforms due to the action of the Lorentz force outside the defect. At the depinning transition, $j = j_{c}$, the Lorentz force is exactly balanced by the maximal pinning force, $f_{p}$, the defect can provide. A current exceeding $j_{c}$ strips the vortex line off the inclusion and the motion of the freed line generates a finite voltage.

For pins with  lateral sizes larger than the coherence length $\xi$, the depinning process is strongly influenced by deformations caused by the Lorentz force on the free vortex segments. These deformations force the vortex entry and exit points to slide along the inclusion surface and approach each other. When the local angle $\theta$ between the segments at these points is reduced below a certain value ($\pi/2$ in isotropic case), the segments start to attract. At a somewhat lower critical angle $\theta = \theta_{c}$, the static configuration becomes unstable, the segments reconnect, and the flux line moves away from the defect. The current $j$ imposing this critical angle defines the critical current $j_{c}$. For this release scenario, the pin-breaking force is mostly determined by the line tension of the vortex rather than by properties of the pin itself.

For a spheroid inclusion with sizes $a_{x} = a_{y} > \xi$ and $a_{z}>\xi_{z}=\xi/\gamma$ in an anisotropic 3D superconductor,  scaling analysis leads to the following form for the pin-breaking force%
\footnote{We note that this result is different from the naive estimate $(\varepsilon_{0} a_z/a_x)\ln(a_x/\xi)$, which is obtained by dividing the pinning energy  $(\varepsilon_{0} a_z)\ln(a_x/\xi)$ by the defect size $a_x$. The reason of this discrepancy is the strong deformation of the vortex line during the depinning process.}
\begin{align}
f_{p}(a_{x},a_{z},\gamma)=\mathsf{G}_{1}\Big(\frac{\gamma a_{z}}{a_{x}}\Big)\frac{\varepsilon_{0}}{\gamma}
\ln\Big[\mathsf{G}_{2}\Big(\frac{\gamma a_{z}}{a_{x}}\Big)\frac{a_{x}}{\xi}\Big],\label{eq:fp-gen}
\end{align}
where $\gamma$ is the anisotropy factor, $\varepsilon_{0}=(\Phi_0/4\pi \lambda)^2$ is the scale for the vortex line energy, $\lambda$ is the London penetration depth, and $\mathsf{G}_{i}$ are dimensionless functions of order unity. 
\begin{figure}[tb]
\centering
\includegraphics[width = .45\textwidth]{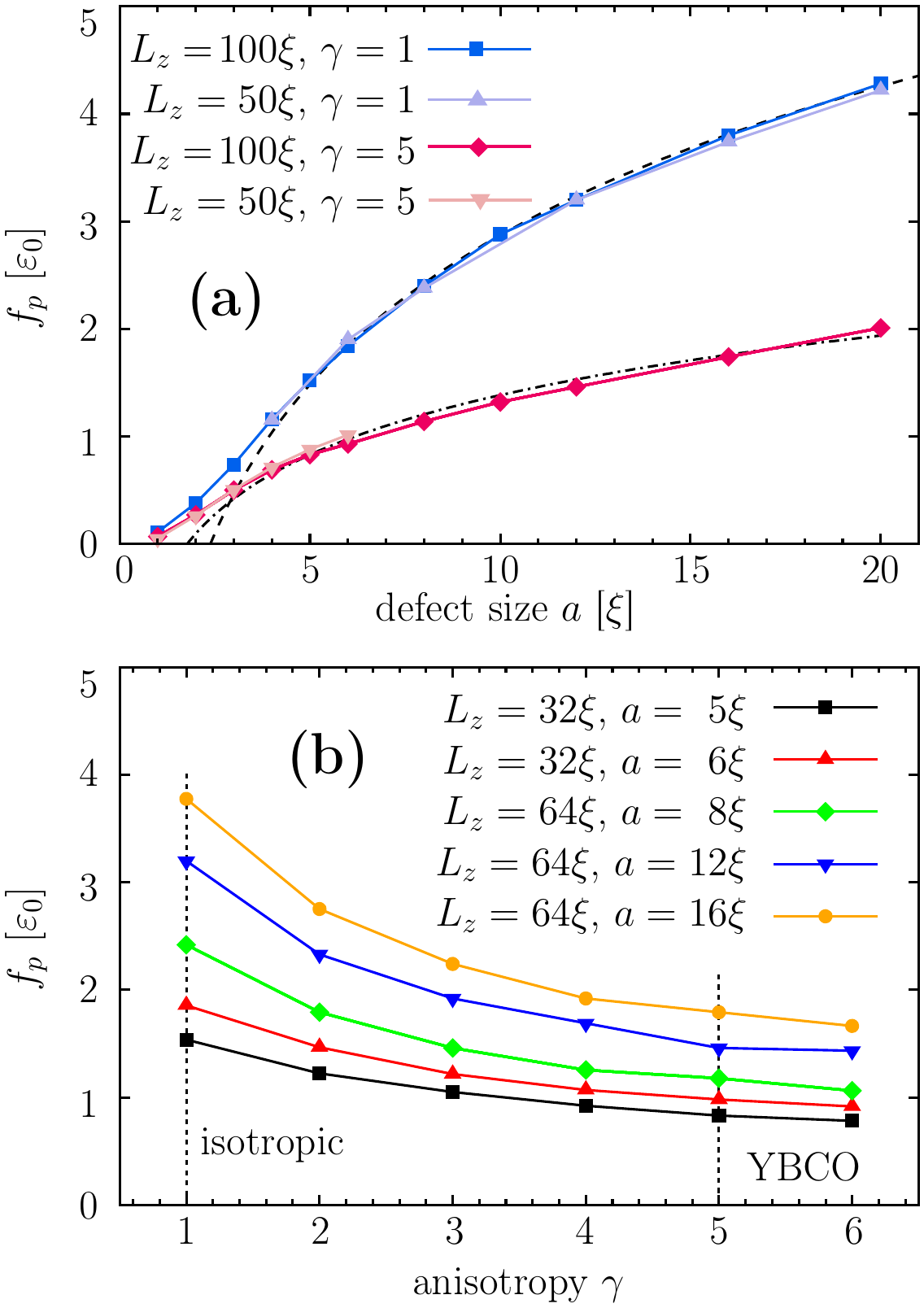}
\caption{
\textbf{(a)}
Pin-breaking force $f_{p}$ necessary to depin a single vortex from a spherical inclusion of diameter $a$. Colors differentiate between anisotropies, i.e. $\gamma = 1$ (blue) and $\gamma = 5$ (red), while the color shade indicates the system size. The dashed and dash-dotted lines are logarithmic fits suggested by \eq \eqref{eq:fp-gen}.
\textbf{(b)}
Dependence of the pin-breaking force on the anisotropy $\gamma$. Colors/symbols are associated with defect sizes.
}
\label{fig:fp}
\end{figure}

We have executed two sets of time-dependent Ginzburg-Landau simulations to study the depinning process from an isolated defect: The first set focuses on the dependence of $f_{p}$ on the defect size (in the range $a\! =\! 2$-$20 \xi$) for both an isotropic material, $\gamma\! =\! 1$, and a superconductor with anisotropy $\gamma\! =\! 5$, corresponding approximately to the anisotropy found in YBa$_2$Cu$_3$O$_{7-x}$. These simulations are done in a cuboid volume with side lengths $L_{x}, L_{y}, L_{z}$, each measuring either $50 \xi$ or $100 \xi$ with $128$ or $256$ mesh points respectively. At the volume's center [defining the origin $(x,y,z) = \vec{0}$ of the coordinate system] we place a spherical inclusion. Inside the inclusion the linear coefficient in the time-dependent Ginzburg-Landau equation assumes a negative value $\epsilon = -1$, see Appendix \ref{Sec:model}. In the second set we investigate the anisotropy dependence of $f_{p}$ in the range $1\!\leq\!\gamma\! \leq \!6$ for fixed inclusion diameters between $5\xi$ and $16\xi$. These simulations are done for a cubic system with lateral sizes $32\xi$ or $64\xi$ and two mesh points per $\xi$.

\begin{figure}[tb]
	\centering	
	\includegraphics[width = .49\textwidth]{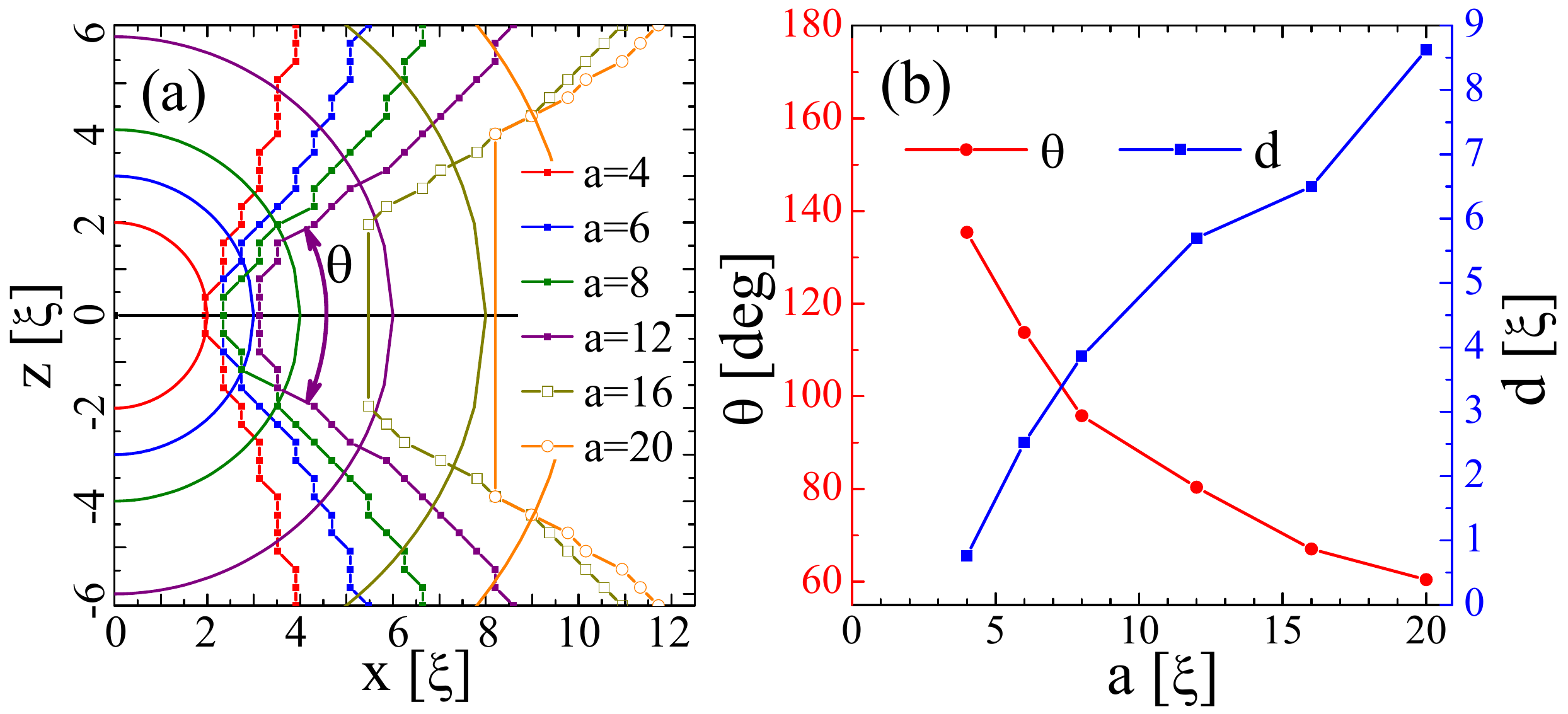}
	\caption{\textbf{(a)} Enlarged view of the critical vortex configuration in the vicinity of particles of different sizes $a$. The solid circle segment lines mark the particle boundaries and the lines with symbols show vortex configurations in the critical state. \textbf{(b)} Critical angle $\theta$ and distance $d$ between vortex tips versus defect diameter.}
	\label{Fig-CritConf}
\end{figure}

Figure \ref{Fig-ParticlePinBreak} illustrates the vortex-line depinning dynamics in an isotropic superconductor  for two spherical inclusions with diameters $4 \xi$ and $16 \xi$, respectively. The left column of Fig.~\ref{Fig-ParticlePinBreak} shows representative time dependences of the current ramping (blue steps) and the resulting space-averaged electric field (red). At the critical current, the flux line detaches from the inclusion, as manifested by a rapid increase of the electric field. The columns on the right of Fig.~\ref{Fig-ParticlePinBreak} show isosurfaces of the order parameter for representative configurations near criticality. These isosurfaces image both the vortex line and the inclusion. Animations of line depinning can be found in the Supplementary Materials\cite{SupplMat-arxiv}.

Once the critical current $j_{c}$  is obtained from the numerical simulations (here, in units of $j_{0} = c\Phi_{0}/8\pi^{2}\lambda^{2}\xi$ comparable to the depairing current $j_{\mathrm{dp}} = 2j_{0}/3\sqrt{3}$), the pin-breaking force can be calculated from the relation $f_{p} = (\Phi_{0}/c)j_{c}L_{z} = 2\varepsilon_{0}(j_{c}/j_{0})(L_z/\xi)$. The size dependence of $f_{p}$ is presented in Fig.~\ref{fig:fp}(a) for two values of the anisotropy, where the force is expressed through the natural scale $\varepsilon_{0}$ of the vortex line tension. We find that for $a>4\xi$, the size dependence of  $f_{p}$ is well described by the logarithmic function, as suggested by Eq.\ \eqref{eq:fp-gen}, see  fits in Fig.\ \ref{fig:fp}. In particular, for the isotropic case, the dependence of the maximal pinning force on the defect size for large inclusions is well described by $f_p(a) \approx 2 \varepsilon_{0}\ln(0.42 a/\xi)$. The rather weak logarithmic dependence on the particle size confirms the assumption that the pin-breaking force is mostly determined by the vortex line tension. The anisotropy dependences of the pin breaking force is shown in Fig.~\ref{fig:fp}(b) for several inclusion sizes. The decrease of $f_{p}$ with the anisotropy factor $\gamma$ indicates again the relevance of the line tension in the depinning process. We find that $f_p(a)\propto \ln (a)$ for all anisotropies and the coefficient in front of the logarithm monotonically decreases with the anisotropy factor but slower than $1/\gamma$. The function $\mathsf{G}_{1}(\beta)$ in Eq.\ \eqref{eq:fp-gen} is well described by the simple dependence, $\mathsf{G}_{1}(\beta)\approx 1.5+0.5\beta$ for $1\!<\!\beta\!<\!6$; note that $\beta\!=\!\gamma$ for spherical particles with $a_z\!=\!a_x$. We also find that $\mathsf{G}_{2}(\gamma)$ is a non-monotonic function weakly-varying  within the range $0.42\text{ - }0.56$, which can be interpolated as $\mathsf{G}_{2}(\gamma)\approx 0.354+0.097\gamma-0.011\gamma^2$.

Investigating the vortex shape, in the vicinity (including the inside) of the inclusion, we have identified two parameters characterizing the last stable configuration: (i) the critical angle between the vortex segments at the points where they enter the inclusion (defined as an angle between the linear interpolations at these points) and (ii) the distance between the entrance points. From the critical vortex-line configurations in the vicinity of spherical defects, see Fig.~\ref{Fig-CritConf}, we find a monotonically decreasing critical angle with increasing particle size saturating near $60^{\circ}$ for the large-size particle. At the same time, the distance between the vortex tips intersecting with the inclusion's surface does not show a saturation behavior. As a consequence, the distance between the vortex segments significantly exceeds the coherence length for large inclusions. Indeed, it reaches up to $9 \xi$ for the largest particle $a = 20 \xi$. We conclude from this analysis that the critical angle---rather than the nearest distance between vortex segments---is the relevant parameter determining the critical state.

From a more technical perspective, we find that the depinning force is generally robust against changing the system size. In the anisotropic case for $a = 5\xi$ and $a = 6\xi$, the critical force in the cubic volume $(50 \xi)^{3}$ slightly deviates (5-10\%) from the larger cubic system $(100 \xi)^{3}$. We attribute this effect to the large vortex deformation $\sim 20$-$30\xi$, i.e., reaching a significant fraction of the lateral system size. For this reason, we limit ourselves to the larger simulation volume for inclusions with $a > 6\xi$.

\section{Interaction of a vortex lattice with an isolated spherical inclusion}\label{sec:manyvortices-singledefect}

\begin{figure}[t]
\centering	
\includegraphics[width = .43\textwidth]{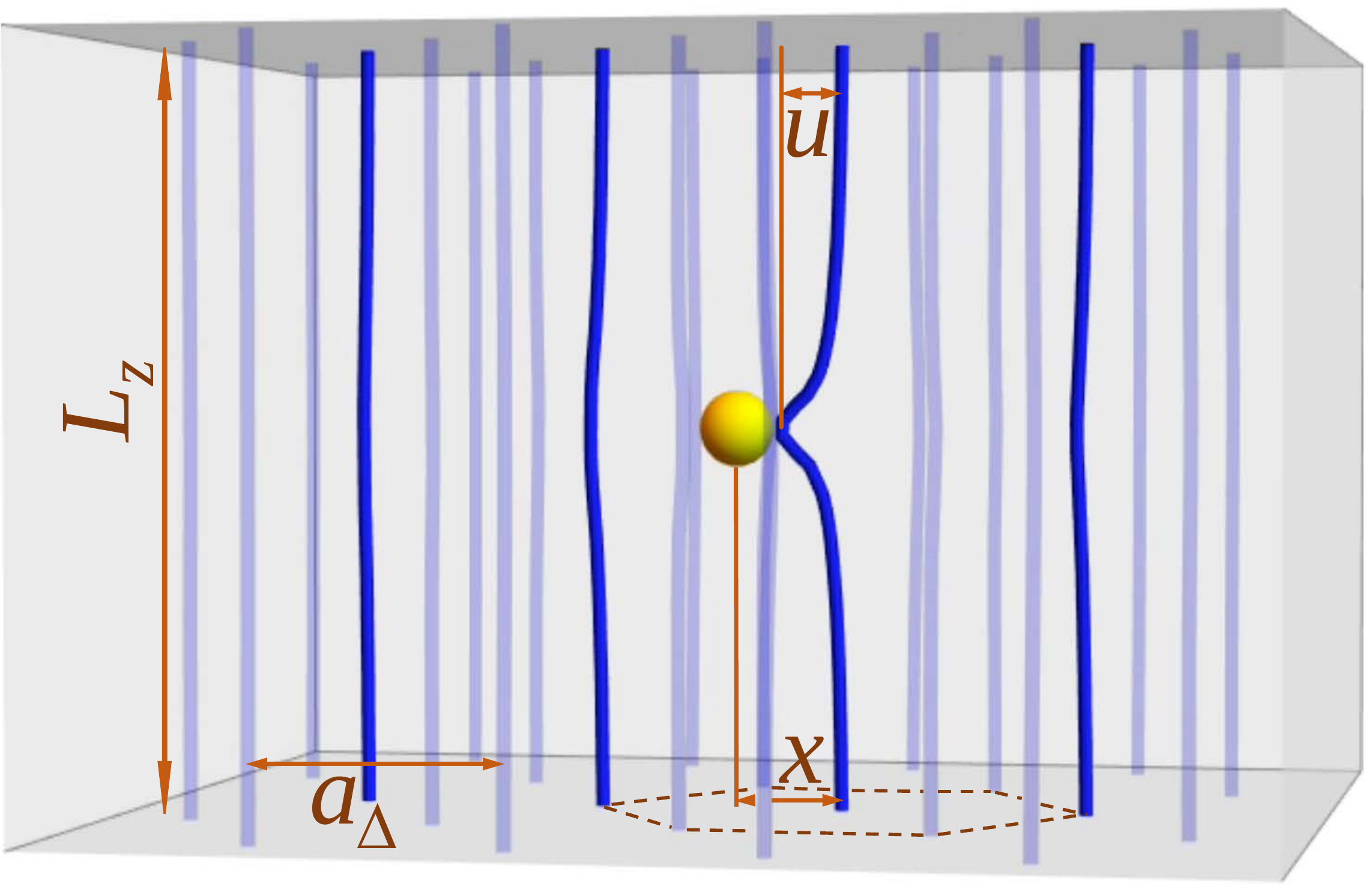}
\caption{An isolated inclusion interacting with the vortex lattice. This simulation snapshot highlights the vortex row (thick, dark blue) impacting the spherical defect in their path. Vortices from lateral rows are shown as well (semi-transparent). The vortex (lattice) displacement $x$, the deformation $u$ imposed on the pinned flux line, and the system's vertical size $L_{z}$ are highlighted.}
\label{fig-InclVortLatt}
\end{figure}
In this section we explore the interaction of an isolated inclusion with a vortex lattice at different magnetic fields. We investigate the simplest geometry of $N_{v}$ vortices with length $L_{z}$ arranged in an ideal lattice with period $a_{\sss \triangle}$, which is driven along a nearest-neighbor direction ($x$ axis) by the current $j$ flowing along the $y$ axis. A spherical inclusion is placed at the origin, withing the vortex-row plane, as illustrated in Fig.~\ref{fig-InclVortLatt}. The pin-breaking force is typically considered as an intrinsic property of a defect. We will see, however, that for large-size inclusions a finite magnetic field strongly affects the depinning process leading to a nontrivial field dependence of the pin-breaking force.  Before studying this problem numerically, we first give a brief account of the theoretical expectations.

\subsection{Theoretical background}
\subsubsection{Linear response}
At small forces, the pinned flux-line lattice is expected to respond linearly to an external force. When a total Lorentz force $f_{L} = N_{v} L_{z} \Phi_{0} j/c$ acts on the vortex system, the defect will react with the same force of opposite sign, i.e., $f_{\mathrm{pin}} = - f_{L}$. At the same time, this force displaces the vortex lattice by a distance $x > 0$ relative to its unperturbed position. Furthermore, a maximal deformation $u < 0$ is imposed on the pinned vortex, see Fig.\ \ref{fig-InclVortLatt}. In the linear regime, where $x \propto f_{\mathrm{pin}}$, this deformation satisfies an elastic force-balance equation of the form
\begin{align}\label{eq:el-force-balace}
   f_{\mathrm{pin}}(x) = \C u.
\end{align}
Here the effective spring constant $\C$ relates to the elastic Green's function $G_{\alpha \beta}(\vec{r},z)$ giving the $\alpha$ component of the vortex lattice displacement at the point $(\vec{r},z)$ caused by the $\beta$ component of the force acting at the origin. Usually, the defect is approximated by a point-like pin having a $\delta$-shaped potential. In this case $\C_{0} = [G_{xx}(\vec{0},0)]^{-1}$ and a proper evaluation\cite{KwokRoPP2016, Willa2018a} in the limit $B\ll H_{c2}$ provides the estimate,
\begin{align}\label{eq:Cbar-of-0}
\C_{0} \approx 3 \sqrt{\eone \varepsilon_{0}}/a_{\sss \triangle} \approx 1.2 (\varepsilon_{0}/\gamma\xi)(B/H_{c2})^{1/2},
\end{align}
where $\eone \approx \varepsilon_{0}/\gamma^2$ is the vortex line tension.

The point-force approximation is however expected to break down once the defect's vertical size $a_{z}$ meets or exceeds the characteristic healing length $\ell_{h} = a_{\sss \triangle}/\gamma$ of the vortex perturbation. As shown in Appendix \ref{sec:lin-el}, the elastic Green's function $G_{xx}(\vec{r},z)$ is then probed along a finite length $\approx a_z$ parallel to the $z$-axis, yielding
\begin{align}\label{eq:Cbar-of-a}
   \C(a_{z}) = \C_{0} \frac{\chi a_{z}/\ell_{h}}{\ln(1 + \chi a_{z}/\ell_{h})},
\end{align}
with $\chi$ a numerical constant of order unity (numerical calculations give $\chi\approx 2-2.5$). For small inclusions $a_{z} \ll \ell_{h}$, the result reduces to the zero-size expression \eqref{eq:Cbar-of-0}. In the opposite limit of long inclusions, $a_{z} \gg \ell_{h}$, the spring constant reduces to $\C_{\sss \mathrm{2D}} \approx \varepsilon_{0}/a_{\sss \triangle}^{2} \ln(a_{z}\gamma/a_{\sss \triangle})$, i.e., the elasticity of vortices in two-dimensional films. The logarithmic divergence has to be properly cut off.

Regarding the applicability of expression \eqref{eq:Cbar-of-a}, it is worth noting that the requirements $a \ll a_{\sss \triangle}$ (low-field, in-plane) and $a_{z} \notll \ell_{h} = a_{\sss \triangle}/\gamma$ (along $z$) are antithetic. For spherical inclusions in an isotropic superconductor ($\gamma = 1$), this only leaves little room for this type of defect-size effect. Therefore, the result \eqref{eq:Cbar-of-a} is mostly relevant for either anisotropic superconductors ($\gamma > 1$) or for elongated inclusions [$a_{x} = a_{y} = a$ and $a_{z} > \gamma a$].

\subsubsection{Pinning force profile}
The linear elastic approximation breaks down at larger forces, i.e., near the depinning transition. More precisely, a vortex at a distance $x$ away from the defect deforms according to the nonlinear equation\cite{BlatterGK:2004}
\begin{align}\label{eq:non-linear-force-balance}
f_{p}(x + u) = \C u
\end{align}
with $f_{p}(r)$ the (bare) force associated with the defect's pinning well. The system described by this  equation goes through a weak-to-strong  pinning transition when the Labusch parameter $\kappa\! \equiv \! \max_{r}[f_{p}'(r)]/\C$ reaches unity (the prime denotes the derivative with respect to the function's argument, here $r$). In the strong-pinning regime, $\kappa\! >\! 1$, the equation \eqref{eq:non-linear-force-balance} exhibits multiple solutions $u(x)$ in a finite interval $x \in [x_{-}, x_{+}]$, the boundaries of which are characterized by a softening of the vortex deformation, $u'(x \to x_{\pm}) \to \infty$.

Similar to the linear-response regime, \eq \eqref{eq:el-force-balace}, a pinning force $f_{\mathrm{pin}}(x) \equiv f_{p}[x + u(x)]$ may be defined. Due to the existence of multiple solutions $u(x)$, the force profile $f_{\mathrm{pin}}(x)$ itself is also multivalued. It is this pinning landscape that is probed by the vortex state as it adiabatically moves through the defect. If initially pinned at $x = 0$, the vortex state assumes a pinned solution until reaching the critical value $x_{+} \approx u_{c}\equiv f_{p}/\C$.  This point defines the \emph{pin-breaking force} $f_{p} \equiv \max_{r}[f_{p}(r)]$, which in this basic picture is a \emph{field-independent} parameter. The vortex deformation in the critical state $u_{c}$ then depends on the field strength $B$ through the elastic spring constant $\C(B)\! \propto\! B^{1/2}$ and is proportional to $a_{\sss \triangle}$. At the point $x\!=\!x_{+}$ the pinned branch terminates, the vortex pinches off from the defect, and only the unpinned solution is realized. On the other hand, as a vortex approaches the defect from the opposite side, the unpinned solution is realized for $x\! <\! -x_{-}$, where it undergoes a trapping instability and snaps into the defect.

The described basic scenario assumed in the majority of theoretical studies \cite{Labusch1969, LarkinO:1979, Vinokur1990, Coffey1991, OvchinnikovI:1991, BlatterFGLV:1994, Brandt:1995, BlatterGK:2004, GurevichSST07, Buchacek2018a} is valid if the critical deformation $u_{c}$ is much smaller than the lattice period $a_{\sss \triangle}$. This condition may break for large inclusions at sufficiently strong magnetic field, especially in strongly-anisotropic superconductors. In this case, the trapping instability of a second vortex approaching the defect may occur \emph{before} the departure of the pinned flux line. The force-balance equation for this second vortex reads\cite{Willa2018a}
\begin{align}\label{eq:second-vortex}
f_{p}(x_{s} + u_{s}) = \C(u_{s} - \Gamma u)
\end{align}
where $u_{s}$ is the deformation of the second vortex at $x_{s} = x - a_{\sss \triangle}$. The term $\Gamma u$ describes the contribution to the second-vortex displacement coming from the force acting on the pinned vortex with  $\Gamma \approx G_{xx}(0)/G_{xx}(a_{\sss \triangle})$ (estimates provide $\Gamma \approx 0.23$, see Ref.\ \sbonlinecite{Willa2018a}). This equation gives a trapping instability of the unpinned vortex at $x_{t} = (a_{\sss \triangle} - x_{-})/(1-\Gamma)$. For $x_{t} < x_{+}$, the depinning scenario changes qualitatively. In this case the second vortex jumps into the inclusion and  forces the departure of an (already) pinned one. This yields the reduction of the pin-breaking force
\begin{align}\label{eq:pinfrc-trap}
f_{p}(B) \approx (x_{t} / x_{+}) f_{p}.
\end{align}
As $x_{+}$ is approximately  proportional to $a_{\sss \triangle}$, the field dependence of the pin-breaking force is dominated by the factor $1\! -\! x_{-}/a_{\sss \triangle}$, hence extrapolating to zero where the intervortex distance $a_{\sss \triangle}$ matches $x_{-}\! \approx a$ at $B \approx \Phi_{0}/a^{2}$. The condition  $x_{t} < x_{+}$ also implies that the inclusion is always occupied.

It is beyond the scope of the above analysis to tell if the trapping of the second vortex causes the pinned vortex to leave. In fact, it is reasonable to assume that for sufficiently large fields, $B \gg \Phi_{0}/a^{2}$, the defect will realize a double-occupied pinning ground state.

\subsubsection{Averaging and observable quantities}
In a realistic scenario of a superconductor with a low density $n_{p}$ of randomly distributed defects, each vortex-to-pin distance $x$ (along the force direction) occurs with equal probability. In addition, the inclusion may be located at a finite impact distance $y$ in the direction transverse to the vortex motion. In this case, the vortex is trapped by the inclusion only if $|y|\! <\! x_{-}$. 
Macroscopic observables (such as the critical current measured in experiments) arise from a proper average over the realized states of the pinning force. For this purpose, the direction transverse to the vortex motion is approximately accounted for by the factor $2x_{-} /a_{\sss \triangle}$, i.e., identifying vortices that impact the defect within a transverse trapping distance $2 x_{-}$ with one impacting head-on the defect ($y = 0$). For the specific case of the critical current, one then finds\cite{BlatterGK:2004, Willa2015a, Willa2016, Willa2018a}
$j_{c}\! \approx \! (c n_{p}/B)(2x_{-}/a_{\sss \triangle}) \langle f_{\mathrm{pin}}\rangle$.
In this regime the critical current decreases with the field approximately as $1/\sqrt{B}$. At high fields, i.e., when $a_{\sss \triangle}\! \approx x_{-}$, the bulk critical current further simplifies to
\begin{align}\label{eq:jc}
   j_{c} &\approx \frac{c n_{p}}{B}\langle f_{\mathrm{pin}}\rangle.
\end{align}
In the simplest case with field-independent $\langle f_{\mathrm{pin}}\rangle$ the critical current is expected to decay as $1/B$.

\subsection{Numerical results}\label{sec:num}
In order to explore the pinning properties in different regimes beyond simple analytical estimates, we investigate the interaction of an isolated inclusion with a vortex lattice using large-scale time-dependent Ginzburg-Landau simulations.\cite{SadovskyyJComp2015} Specifically, we initialized the simulations with an ideal flux-line lattice composed of $N_{v} = 36$ vortices, hence adjusting the system's lateral dimensions $L_{x} = (3/4)^{1/2} L_{y} = 6 a_{\sss \triangle}$ to the field strength $B = (4/3)^{1/2} \Phi_{0}/a_{\sss \triangle}^{2}$. For the vertical system size, we used $L_{z} = 50\xi$. Furthermore, a central row of vortices was aligned along $y = 0$. This configuration allows to probe the linear-elastic (see Appendix \ref{sec:lin-el}), pin-breaking, and dynamic regimes numerically by placing defects of various diameters $a$ at the volume's origin and ramping up the applied current.

\subsubsection{Pin-breaking force}\label{sec:pin-breaking}
\begin{figure}[tb]
\centering
\includegraphics[width = .47\textwidth]{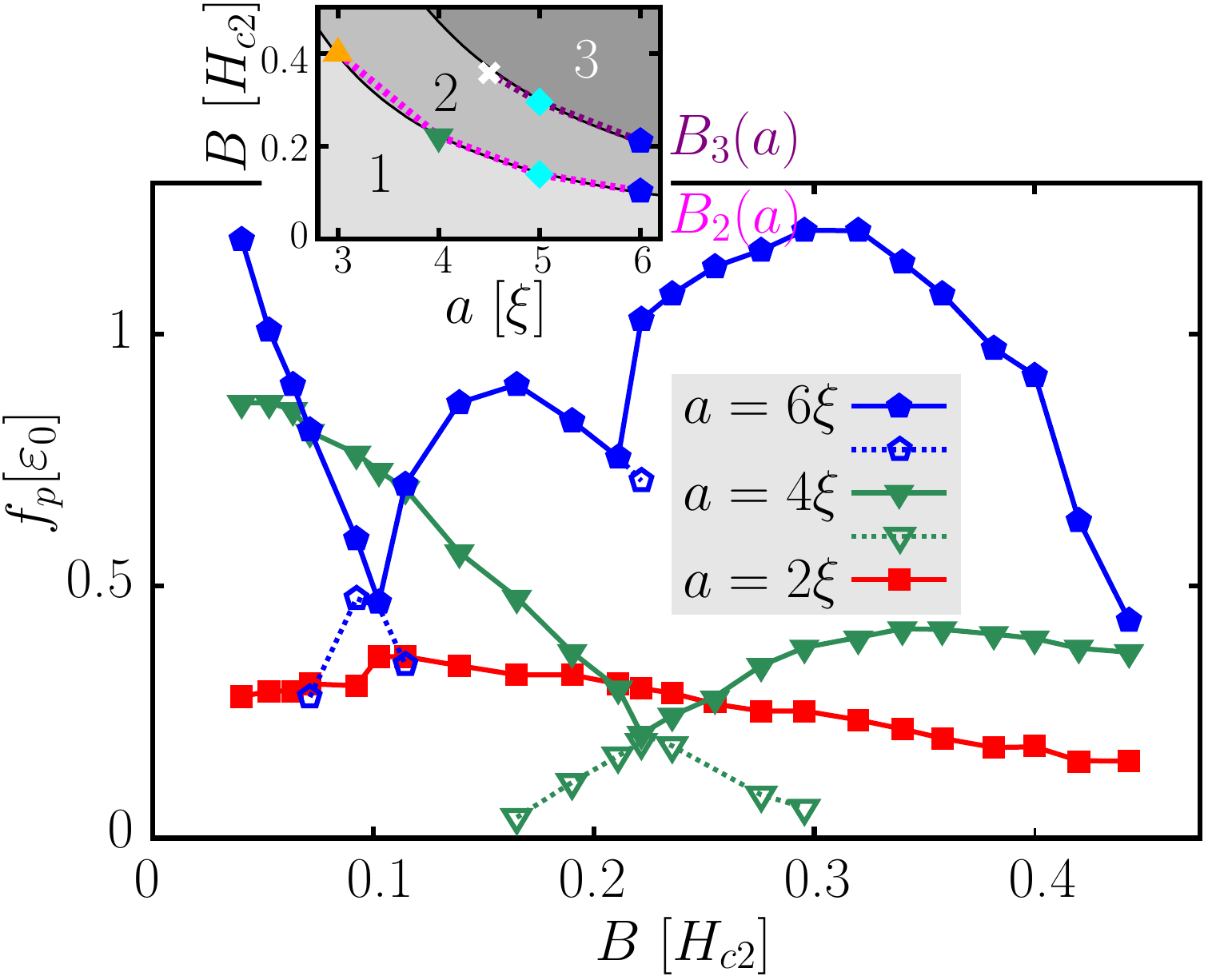}
\caption{Field-dependent pin-breaking force $f_{p} = N_{v} L_{z} \Phi_{0} j_{c}/c$ for three different defect sizes $a = 2\xi$ (red, square), $4\xi$ (green, triangle), and $6\xi$ (blue, pentagon). The transitions between different occupation states are seen as sharp cusplike dips of $f_{p}(B)$ for $a = 4\xi$ and $6\xi$. In addition to the \emph{absolute} maximum forces shown by solid symbols, for $a = 4\xi$ and $6\xi$, we also show the \emph{local} maxima $f_{p}^{\mathrm{meta}}$ for metastable states extracted from dynamics simulations by open symbols (dashed), see text. The transition fields for single-to-double [$B_2(a)$, magenta line] and double-to-triple [$B_3(a)$, purple line] occupancy---shown in the inset---follow the empirical laws $B_{2,3}(a)/H_{c2} = \mathrm{C}_{2,3} (\xi/a)^{2}$ (black solid lines), with $\mathrm{C}_{2} = 3.5$ and $\mathrm{C}_{3} = 7.5$.
}
\label{fig:fp-2D}
\end{figure}

At the defect's pin-breaking force $f_{p}$ the vortex lattice departs from its static configuration and the transition to a dynamic regime is manifested by the onset of dissipation. The appearance of a finite voltage hence provides an ideal criterion to determine the defect's maximal pinning force, $f_p$. We systematically evaluate this key parameter for different inclusion sizes and magnetic fields. The overall behavior of $f_p$ for all studied sizes and fields is illustrated by the waterfall plot in Fig.~\ref{fig:fp-of-a-B}. For a more detailed presentation and quantitative comparison, we show  the magnetic field dependences of the pin-breaking force for three  selected inclusion sizes, $2\xi$, $4\xi$, and $6\xi$ in Fig.~\ref{fig:fp-2D}. Starting with the smallest inclusion, $a = 2\xi$, the pin-breaking force shows two regimes: At low fields, $f_{p}$ slightly increases with increasing magnetic field. We attribute this weak field dependence to the confining effect of the vortex lattice, leading to a rectification of the pinned vortex and an increase of the angle between two vortex pieces entering the inclusion. As a consequence, the critical angle $\theta_{c}$ at which the depinning instability develops is reached at larger forces. At larger fields, $B \gtrsim 0.1 H_{c2}$, the pin-breaking force gradually diminishes, reaching only a fraction of its maximal value near $0.5 H_{c2}$. We attribute this effect to the influence of an unpinned vortex approaching a defect that still traps a flux line.  Nevertheless, for small inclusion in a wide field range, the general scenario follows simple expectations. 

For a larger inclusion, $a > 2\xi$, the field dependence of $f_{p}$ is strongly influenced by the possibility of a (stable) double-occupancy of the defect. Above a certain field $B_2(a)$, the critical state changes abruptly from a single-occupied to a double-occupied defect.%
\footnote{This transition field of the critical-state modification is very close---yet not identical---to the static transition field at which the zero-current ground state changes from single to double occupation of the inclusion.}
Prior to that transition, the pin-breaking force rapidly decreases with field. In this range, the critical state corresponds to trapping of the second vortex which immediately expels the pinned vortex, in agreement with the above consideration, see Eq.\ \eqref{eq:pinfrc-trap}. Above the transition, $B > B_2(a)$, the double-occupied state is stable and the pin-breaking force $f_{p}$ measures its criticality. Due to different nature of the critical state above and below $B_2$, $f_{p}(B)$ features a kink at the transition. Moreover, the degeneracy between the two occupation states at the transition point leads to the pronounced minimum of $f_{p}(B)$.
Above the kink, the pin-breaking force rapidly increases with $B$ due to further stabilization of the double-occupied state. The transition field $B_2(a)$ follows the empirical law $B_2(a) \propto a^{-2}$, i.e., $a_{\sss \triangle}(B_2)/a = \mathrm{const.}$, see the inset in Fig.\ \ref{fig:fp-2D}.
At even higher fields, the competition with other unpinned vortices becomes relevant and the pin-breaking force $f_{p}(B)$ starts decreasing again. The peak effect repeats itself when new vortices are accommodated in the defect, as observed here for large inclusions $a \geq 5\xi$, where triple-occupancy occurs above $B_3$. Whereas at low fields, only the central vortex row (impacting the defect) is involved in the pinning-depinning process, the regime of triple-occupancy typically involves vortices from neighboring vortex rows.

\subsubsection{Quasi-static pinning force profile}\label{sec:dynamic}
\begin{figure}[t!bh]
\centering	
\includegraphics[width = .45\textwidth]{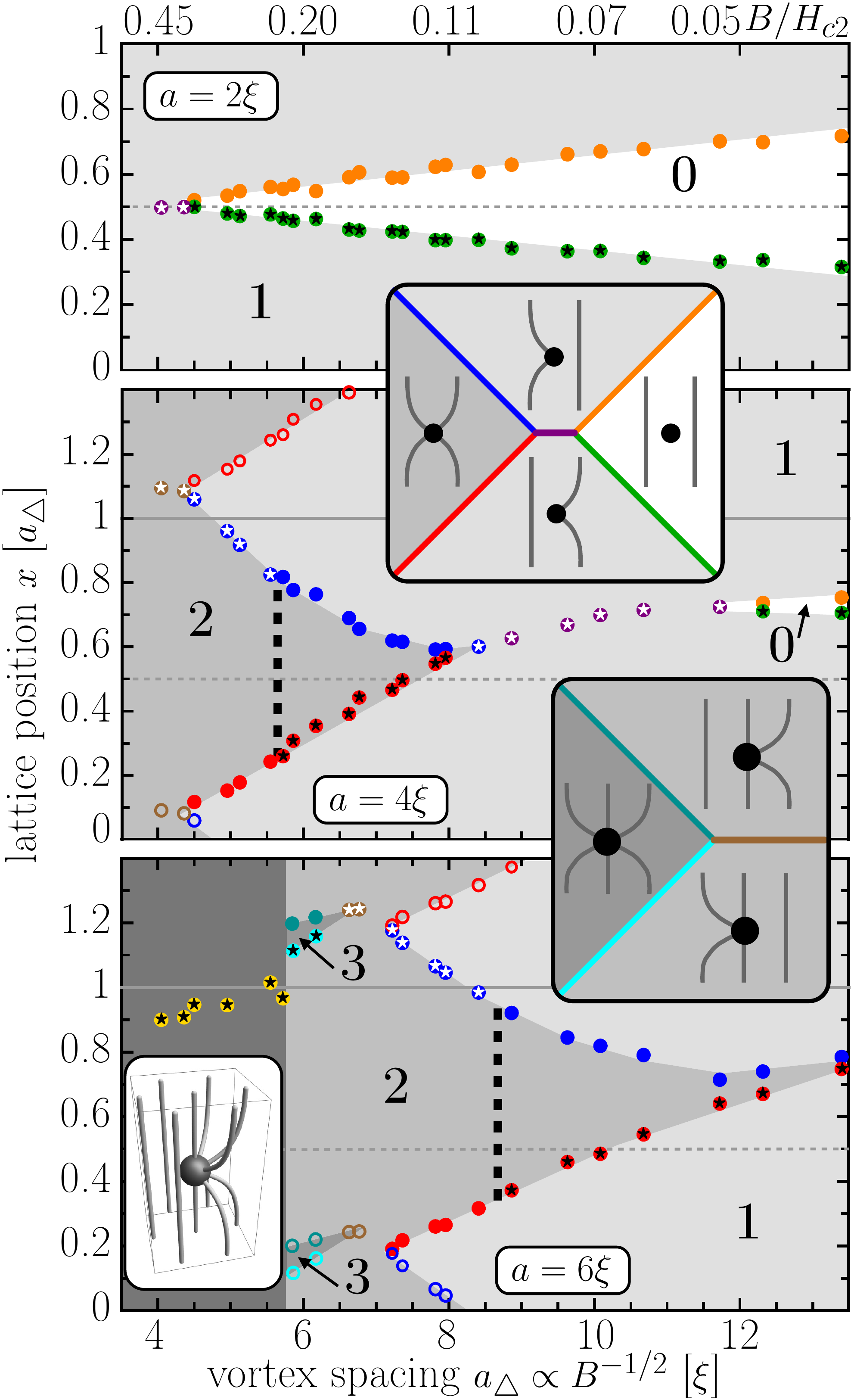}
\caption{Pinning states and boundaries between them for an isolated inclusion of size $a = 2\xi$ (top), $4\xi$ (center), and $6\xi$ (bottom), as obtained from the quasistatic simulations. As the flux lattice slowly moves, a vortex may get trapped by or depin from the inclusion. In the simplest scenario realized for  $2\xi$ and $4\xi$, the defect captures only one defect at the time (green/orange). For higher fields/larger defects, trapping of a free vortex and freeing of the pinned one can occur simultaneously (purple). For $4\xi$ and $6\xi$ at even higher fields, the freeing/trapping process occurs in reversed chronological order (red/blue) and results in a double-occupancy of the defect. For $a=6\xi$ phase boundaries involving three or more vortices (including those from neighboring vortex rows) lead to a rich phase pattern at very high fields. The critical state is marked by a five-pointed star (black or white) and shows a first-order transition at $B_2(a)$ (dashed line). This transition coincides with the kink in $f_{p}(B)$, see Figs.\ \ref{fig:fp-of-a-B} and \ref{fig:fp-2D}. The inset sketches depict the vortex-defect configuration for each region; the colors of the transition lines correspond to those in the main figure.}
\label{fig:phase-diag}
\end{figure}
In order to extract the defect's pinning characteristics beyond the pin-breaking force $f_{p}$ such as the full force profile $f_{\mathrm{pin}}(x)$, we proceed with dynamic simulations at currents slightly larger than the critical current. As the vortex lattice slowly moves, we observe different occupation states of the defect which are shown in Fig.\ \ref{fig:phase-diag} for three inclusion sizes, see also animations in Supplementary Materials \cite{SupplMat-arxiv}.

Small defects ($a = 2\xi$) follow the simplest scheme where the defect is either occupied by one vortex, or empty. Periodically (i.e., as the vortex lattice moves by one lattice period $a_{\sss \triangle}$), one vortex undergoes a depinning transition $1\!\rightarrow\! 0$, while the next vortex gets trapped upon approaching the defect, $0\!\rightarrow\! 1$. Above a certain field, $\sim\! 0.44H_{c2}$, these two transitions become indistinguishable, i.e., one pinned vortex is instantaneously replaced by the next one, $1\!\rightarrow\! 1$. At this point the defect's unoccupied state vanishes. 

For larger defects, $a > 2\xi$, the behavior is much richer. For $a=4\xi$ the conventional scheme $1\!\rightarrow\! 0/0\!\rightarrow\! 1$ is realized only for small fields, $< 0.052 H_{c2}$, and the regime $1\!\rightarrow\! 1$ with simultaneous trapping and depinning occupies an extended field range $0.052<B/H_{c2}<0.11$.
At higher fields an intermediate double-occupied state develops, when the second vortex snaps into the defect before the first one leaves, $1\!\rightarrow\! 2$. This \emph{dynamic} double-occupied phase appears before the double-occupancy becomes the critical state, i.e., already when $B < B_2$. In this intermediate range the system is characterized by two critical forces: (i) to trap the second vortex to the single-occupied pin,  $1\!\rightarrow\! 2$, and (ii) to release one vortex from the double-occupied pin, $2\!\rightarrow\! 1$. Both these forces are plotted in Fig.\ \ref{fig:fp-2D}.
In the wide range of fields $0.052\!<\!B/H_{c2}\!<\!B_{2}/H_{c2}\!\approx\! 0.22$ the maximum pinning force is determined by the trapping instability of an unpinned vortex by the already occupied defect, rather than by the release of the pinned vortex. This causes the field-dependence and rapid decrease of $f_p(B)$ within this range, see Fig.\ \ref{fig:fp-2D}.
For $B>B_2$ the critical state switches to that of a double-occupied defect and $f_p(B)$ starts to increase. At very high fields, $B>0.4H_{c2}$ the system is on the verge of triple-occupancy, causing $f_p$ to decrease again (with increasing field). Above the field $0.44H_{c2}$ trapping of the third vortex is accompanied by the simultaneous release of the first one, $2\!\rightarrow\! 2$, marking the disappearance of the single-occupied state.

For the largest inclusion studied here ($a=6\xi$), the intermediate double-occupied state is observed starting from the lowest field, and the transition to the double-occupation criticality occurs in the same way as for smaller inclusion but at a lower field $B_2 = 0.1H_{c2}$. With further field increase, the intermediate \emph{triple} occupied state appears above $0.17 H_{c2}$ when the inclusion already having two vortices grabs the third one within some range of $x$, $2\!\rightarrow\! 3$. Such state is also observed for a smaller defect with $a = 5\xi$. The change of criticality at a somewhat higher field $B_3 \approx 0.22 H_{c2}$ occurs in a rather complicated way. In fact, the intervortex distance in this field range is already comparable with the inclusion size, causing the critical state to involve trapping of vortices from lateral vortex rows.

\begin{figure}[tb]
\centering	
\includegraphics[width = .48\textwidth]{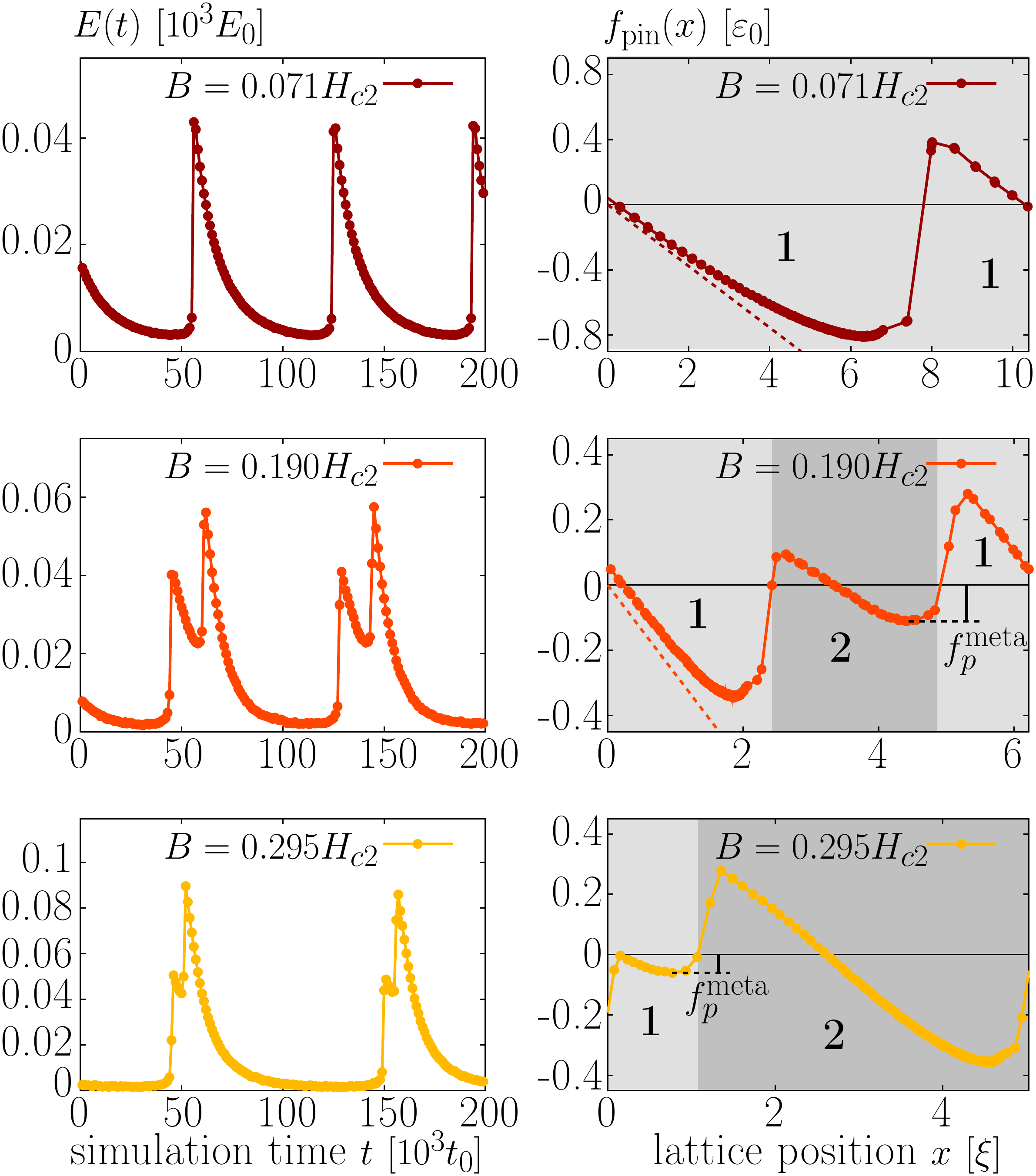}
\caption{Pinning force profiles (right) extracted from the voltage evolution (left) for a single defect of size $a = 4\xi$ at different fields $B/H_{c2} = 0.071,\ 0.190,\ 0.295$. The horizontal axes on the right panels range over one lattice period $[0,a_{\sss \triangle}(B)]$. The small mismatch at $x = 0$ [where $f_{\mathrm{pin}}=0$] is caused by  dynamic effects. The dashed lines indicate the expected elastic response extracted from the static simulations at small forces described in Appendix \ref{sec:lin-el}. At low fields, the defect is occupied by one vortex (light gray). In an intermediate field range the defect is doubly occupied in an extended region (gray). The extremal force $\max_{x}[-f_{\mathrm{pin}}(x)]$ agrees with the pin-breaking force $f_{p}$ obtained from static simulations. Furthermore, the secondary extremum [near $x = 4.5\xi$ ($x = \xi$) for $B/H_{c2} = 0.19$ ($B/H_{c2} = 0.295$)] defines a `critical force' $f_{p}^{\mathrm{meta}}$ for the metastable configuration, see Fig.\ \ref{fig:fp-2D}.
}
\label{fig:fpin}
\end{figure}

In the dynamic regime, we can infer the pinning force profile and extract quantitative defect properties by analyzing the time dependence of the electric field. The lattice motion is governed by the force-balance equation
\begin{align}\label{eq:dynamic-force-balance}
 f_{\mathrm{pin}} = -N_{v} (f_{L} - \eta \bar{v})
\end{align}
where $f_{\mathrm{pin}}$ is the \emph{total} pinning force acting on the vortex lattice, $f_{L} = (\Phi_{0}L_{z}/c)j$ the Lorentz force acting on each vortex along its entire length, and $\eta \bar{v}$ the viscous force of a vortex moving with an average velocity $\bar{v}$. The electric field $E$ recorded in the simulations is related to $\bar{v}$ via
\begin{align}\label{eq:visc-force}
   \eta \bar{v} = \frac{\Phi_{0} E}{c \rho_{\!f\!\!f}} L_{z},
\end{align}
where $\rho_{\!f\!\!f}$ denotes the flux-flow resistivity. From separate simulations\cite{KoshelevPRB16} we have $\rho_{\!f\!\!f} = 1.689 (B/H_{c2}) \rho_{n}$. Combining \eqs \eqref{eq:dynamic-force-balance} and \eqref{eq:visc-force}, we can monitor $f_{\mathrm{pin}}(t) = (\Phi_{0}/c)(j - E/\rho_{\!f\!\!f})N_{v}L_{z}$. In addition, we can extract the vortex-lattice coordinate $x(t)$ from the order-parameter snapshots. Note that $x(t)$ is only defined modulo $a_{\sss \triangle}$. We specifically define $x$ as the (asymptotic, $z \to \pm L_{z}/2$) position of the vortex closest to depinning. For very strong defects, $x$ may therefore exceed one lattice period, see Fig.\ \ref{fig:phase-diag}. The combination of the time-dependent quantities $f_{\mathrm{pin}}(t)$ and $x(t)$ parametrizes the position-dependent pinning force $f_{\mathrm{pin}}(x)$. This procedure relies on the assumption that the velocity is sufficiently small to give the lattice enough time to adjust to its static configuration with the fixed coordinate $x(t)$. Generally, such quasistatic approximation improves with increasing system size.  

Figure \ref{fig:fpin} shows the bare simulation data $E(t)$ (left) and the extracted pinning-force profiles $f_{\mathrm{pin}}(x)$ (right) for defects of size $a\! =\! 4\xi$ and at fields $B/H_{c2}\! = 0.071$, $0.190$, and $0.295$. The first field (upper row) corresponds to  $1\!\rightarrow\! 1$ scenario and the sharp peak in the $E(t)$ dependence marks the simultaneous trapping of arriving vortex and release of the pinned one. The double-peak structure of $E(t)$ for the two other fields is a consequence of the stable double occupied state  for $1\!\rightarrow\! 2/2\!\rightarrow\! 1$ scenario.
The second field is in the range $B < B_2$, i.e., the double occupation is a metastable state, and the event $1\!\rightarrow\! 2$ gives the maximum pinning force. The third field exceeds $B_2$, where the double occupation is a ground state and the maximum pinning force is due to the event $2\!\rightarrow\! 1$. The absolute value of the pinning force is characterized by two maxima corresponding to the transitions $1\!\rightarrow\! 2$ and $2\!\rightarrow\! 1$. The smaller value corresponds to the pin-breaking force from the metastable state, $f_{p}^{\mathrm{meta}}(B)$, e.g., the $2\!\rightarrow\! 1$ transition for $B = 0.190 H_{c2}$. This quantity is highlighted in Fig.\ \ref{fig:fpin}, and the metastable force branches are shown as open symbols in Fig.\ \ref{fig:fp-2D}.

\begin{figure}[tb]
\centering	
\includegraphics[width = .43\textwidth]{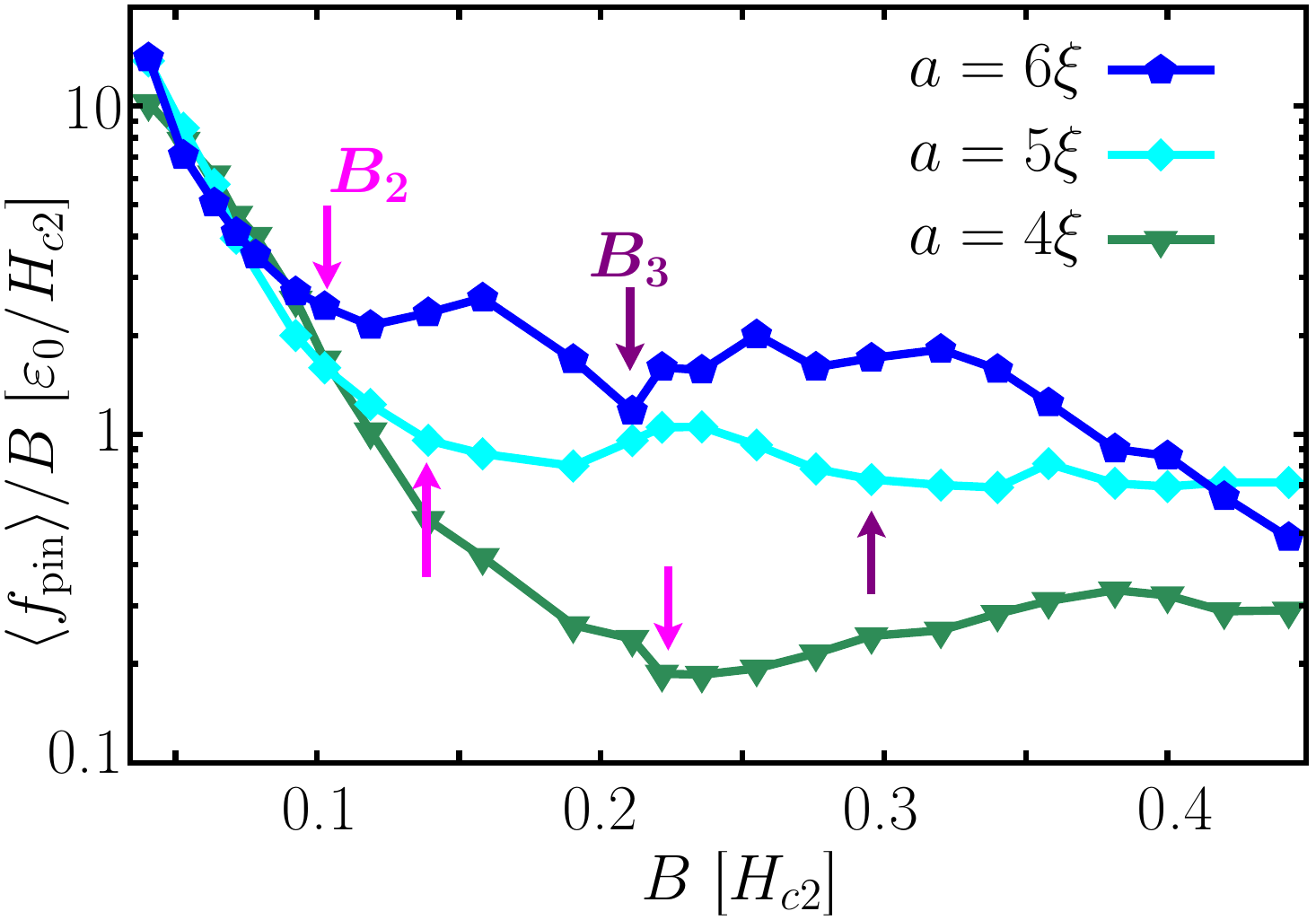}
\caption{Field dependence of the parameter  $\langle f_{\mathrm{pin}} \rangle/B$ for inclusion sizes $a = 4\xi$, $5\xi$, and $6\xi$. This quantity mimics the behavior of the critical current $j_{c}$ for a system with a low density $n_{p}$ of defects, see \eq \eqref{eq:jc}. Despite the averaging, the novel peak effect---caused by multiple occupancy of large defects---remains visible for a realistic system. Arrows indicate the transition from single- to double- ($B_2$) and from double- to triple-occupancy ($B_3$). Notice the logarithmic scale.}
\label{fig:jc-of-B}
\end{figure}
The full pinning force $f_{\mathrm{pin}}(x)$ is the central microscopic ingredient for strong-pinning theory and only its accurate knowledge allows for quantitative predictions. Based on the obtained results for $f_{\mathrm{pin}}(x)$, we can numerically evaluate the average $\langle f_{\mathrm{pin}} \rangle$. In particular, at high magnetic fields, when all defects are occupied, the critical currant is proportional the ratio $\langle f_{\mathrm{pin}} \rangle/B$, \eq \eqref{eq:jc}. Figure \ \ref{fig:jc-of-B} shows the field dependence of this ratio for three defect sizes $4\xi$, $5\xi$, and $6\xi$. Despite some smearing due to the force averaging and despite the overall $1/B$-decay, this quantity still has a nonmonotonic field dependence. The dip and maximum, however, are much less pronounced in comparison to the pin-breaking force. This non-monotonic behavior means that the critical-current in a superconductor with small density of \emph{identical} strong  defects should display a peak effect. The appearance of a maximum in $j_c(B)$ at $B \sim \Phi_{0}/a^{2}$ distinguishes this phenomenon from the `classical' peak effect\cite{LarkinO:1979} due to softening of the line tension near $H_{c2} = \Phi_{0}/2\pi \xi^{2}$. Such a multiple-occupation peak in $j_c(B)$ has been indeed observed in large-scale bulk simulations of the system containing a small density of randomly-located defects.\cite{Willa2018a}

\section{Conclusions}
In this article, we have studied the basic ---yet not satisfactorily addressed before---problem of vortex pinning properties of a single inclusion using time-dependent Ginzburg-Landau simulations. For an isolated vortex, the pin-breaking force scales logarithmically with the defect's size and decreases with increasing anisotropy. The critical state is characterized by the angle $\theta = \theta_{c}$ enclosing the pinned vortex segments, rather than the distance $d$ between these segments.

We find that, contrary to common expectations, the pin-breaking force for a vortex lattice has strong  and nonmonotonic dependence on the field strength. We identify the competition between a single-occupied vs. multiple-occupied defect as the central cause for this field dependence. The transitions between different occupation states are manifested as pronounced cusplike dips in the pin-breaking force and it has shallow maximum in between the neighboring transitions. Proceeding with quasistatic simulations $j/j_{c} - 1 \ll 1$, we have established a route to characterize the microscopic pinning force profile $f_{\mathrm{pin}}(x)$ as a function of the vortex lattice's center-of-mass coordinate $x$. Combining the simulations with results from strong-pinning theory, we have bridged the gap between the microscopic properties of a single inclusion and macroscopic observables [critical current $j_{c}$ and linear $ac$ (Campbell) penetration length] of bulk superconductors. This missing link closes the loop where theory, simulations, and experiments can be compared on a quantitative level.

\begin{acknowledgments}
The authors thank V.\ B.\ Geshkenbein for fruitful discussions. This work was supported by the U.S. Department of Energy, Office of Science, Materials Sciences and Engineering Division.
The TDGL code was developed within the Scientific Discovery through Advanced Computing (SciDAC) program funded by U.S.\ Department of Energy, Office of Science, Advanced Scientific Computing Research and Basic Energy Science.
R.\ W.\ acknowledges funding support from the Early Postdoc. Mobility fellowship of the Swiss National Science Foundation.
\end{acknowledgments}

\appendix

\section{Numerical routine}
\label{Sec:model}
The time-dependent Ginzburg-Landau (TDGL) theory provides a simple, yet sufficiently accurate framework for describing the slow dynamics of a superconductor near the depinning transition. The temporal evolution of the order parameter $\psi$ is thereby governed by the TDGL equation
\begin{align}\label{eq:TDGL}
\mathfrak{u}(\partial_t + i \mu)\psi = \epsilon(\boldsymbol{r})\psi &- |\psi|^2\psi\\ \nonumber
         &+\!\! \sum_{k=x,y,z} \!\! \tilde{\xi}_k^{\,2}(\nabla_k - i A_k)^2\psi + \zeta(\vec{r},t),
\end{align}
written here in its dimensionless form. Time and distance are measured in units of $t_{0} = 4\pi \lambda^{2} / \rho_{n} c^{2}$ and the in-plane coherence length $\xi$ respectively, with $\lambda$ the in-plane penetration depth, $\rho_{n}$ the normal-state resistivity, and $c$ the light velocity. The reduced relaxation rate $\mathfrak{u}$ controls the system's evolution in time. The scalar ($\mu$) and vector ($A_{k}$) potentials enter in a gauge-invariant form and the $\delta$-correlated Langevin force $\zeta(\vec{r},t)$ introduces thermal noise to the system. A uniaxial mass anisotropy $\gamma$ is introduced by rescaling the (dimensionless) coherence lengths $\tilde{\xi}_{k}$ to $\tilde{\xi}_{x} = \tilde{\xi}_{y} = \gamma \tilde{\xi}_{z} = 1$. Finally, the function $\epsilon(\vec{r})$ allows to control the local critical temperature, and hence is suitable for modeling pinscapes. In the infinite-$\lambda$ approximation, the vector potential takes the simple form $\vec{A} = [0, (B/H_{c2})x, 0]$, with $B$ the magnetic field strength along $z$ (crystallographic $c$ axis), $H_{c2} = \Phi_{0}/(2\pi \xi^{2})$ the upper critical field. A uniform electrical current $j$ (measured in units of $j_{0} = c\Phi_{0}/8\pi^{2}\lambda^{2}\xi$) applied along the $y$ direction, will act with Lorentz force $\Phi_{0}j/c$ (per unit length along $x$) on the flux line. In these units, the depairing current reads $j_{\mathrm{dp}} = (2/3\sqrt{3})j_{0} \approx 0.385 j_{0}$. The motion of flux lines is associated with a finite electric field $E = -\partial_{t} A_{y} - \nabla_{y} \mu$ (generated along $y$) and measured in units of $E_{0} = \xi H_{c2}/c t_{0}$.

In order to solve the TDGL equation numerically for relatively large three-dimensional systems, e.g., $100\xi$ in all three dimensions, we use a parallel iterative solver, implemented for graphics processing units (GPU). Implementation details and benchmark analyses of this routine are published elsewhere.\cite{SadovskyyJComp2015} Note, that the discreteness of the numerical mesh (typically $h_{z} = \xi/2.56$), naturally models a layered superconductor when the out-of-plane coherence length $\xi_{z} = \xi/\gamma$ drops far below $h_{z}$. The numerical artifact of discretizing the in-plane directions $h_{x} = h_{y} = \xi/2.56$ produces no measurable effect, suggesting a proper continuum limit $h_{x,y} \ll \xi$. All simulations are performed with periodic boundary conditions along $x$ and $y$, and open boundaries along $z$.

\section{Linear elasticity}\label{sec:lin-el}
\begin{figure}[tbh]
\centering	
\includegraphics[width = .48\textwidth]{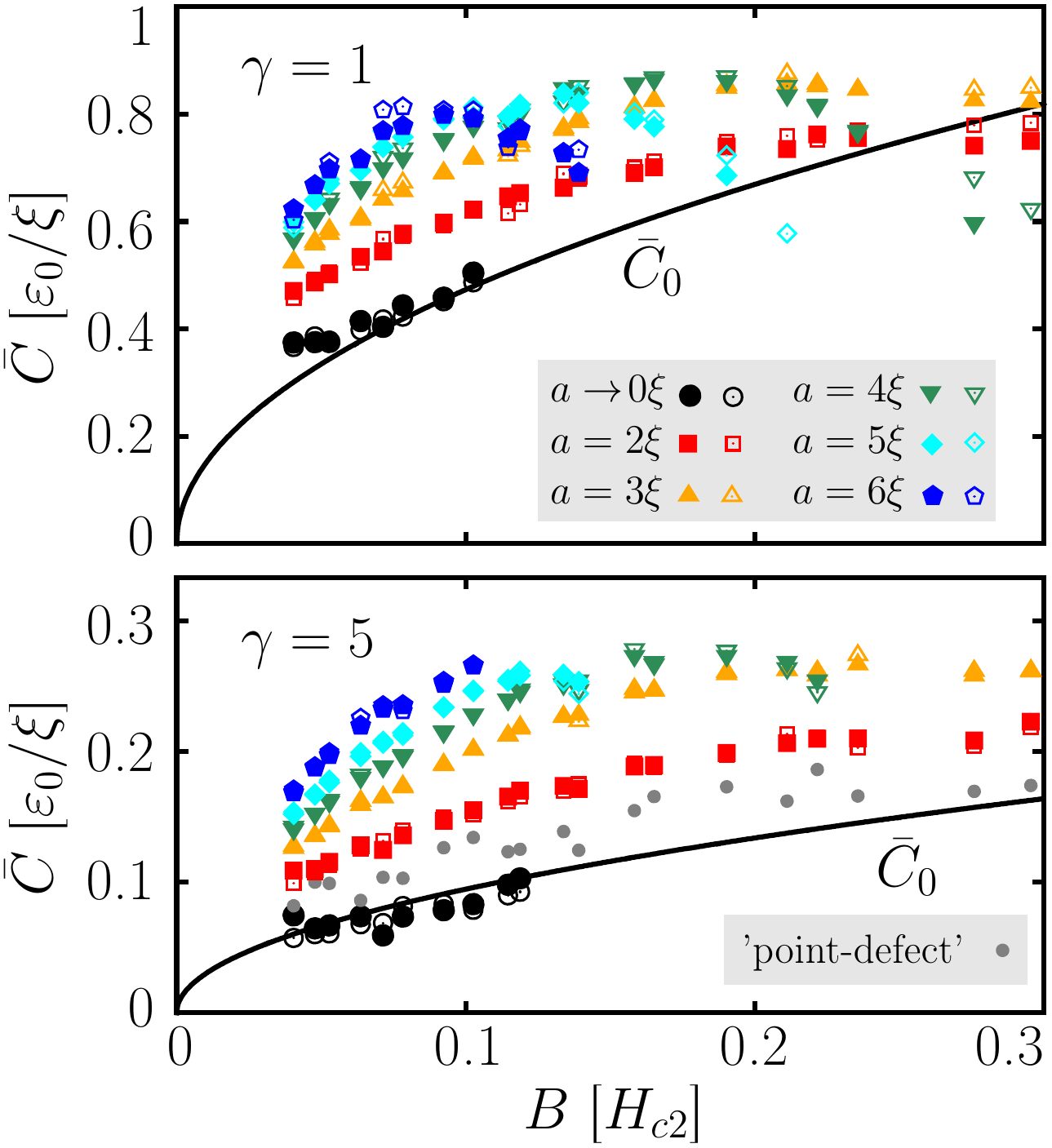}
\caption{The dependences of the effective linear spring constant $\C = f_{L}/u$ on defect size and magnetic field for an isotropic ($\gamma = 1$, top) and an anisotropic ($\gamma = 5$, bottom) superconductor. The elasticity is obtained from numerical simulations by evaluating the vortex displacement $u$ when the system is driven by a fixed (subcritical) force $f_{L} \propto j$, ideally with $j \ll j_{c}$. The results obtained for two current values (full and outline symbols) [these two current values differ by a factor $2$ for $\gamma=1$, and a factor $3$ for $\gamma=5$] confirm the linearity of the response. In each case, $\C_{0}(B)$ may be extracted from fitting the data to \eq \eqref{eq:app-Cbar-of-a}. For the anisotropic case, gray dots indicate simulations for a disc-shaped defect of lateral size $a_{x} = a_{y} = 2\xi$ and a vertical extent of one mesh size, i.e., $a_{z} = h_{z} = \xi/2.56$. Although not spherical, this is the closest one may get to a defect acting as a `point-defect.'}
\label{fig:Cbar}
\end{figure}
We shall derive in this appendix a generalized expression of the spring constant for a finite-sized defect, and study the linear response numerically. If a pinning center acts on the flux line with a distributed force, the spring constant may be modeled by
\begin{align}\label{}
   \C^{-1}(a_{z}) \approx \int_{-\infty}^{\infty} dz\, G_{xx}(\vec{0},z) w(z),
\end{align}
where the elastic Green's function is probed along a finite height $\approx a_z$ along the $z$-axis; the characteristic function $w(z)$ is symmetric [$w(z) = w(-z)$], decays over the length $a_{z}$, and is normalized $\int dz w(z) = 1$. We make use of the convolution theorem to write%
\footnote{Note that the first argument in 
$\hat{G}_{xx}(\vec{0},k_{z})$ is the real-space coordinate,	$(x,y) = \vec{0} \equiv (0,0)$.
}
\begin{align}\label{eq:C-Convolution}
   \C^{-1}(a_{z}) = \int_{-\infty}^{\infty} \frac{dk_{z}}{2\pi} \hat{G}_{xx}(\vec{0},k_{z}) \hat{w}(k_{z}).
\end{align}
Here the hat denotes the Fourier transform over $z$ defined through $\hat{f}(k_{z}) \equiv \int dz f(z) e^{ik_{z}z}$. Whereas the specific shape of $w(z)$ determines the final numerical result, the qualitative behavior is obtained by choosing the simple form $w(z) = e^{-|z|/a_{z}}$. To evaluate the above expression, we follow an integration procedure described in Ref.\ \sbonlinecite{Willa2016}, which we briefly sketch here: The weighted integration of the momentum-space elastic Green's function \cite{LarkinO:1979, Brandt1977a, Brandt1977b}
\begin{align}\label{}
   \hat{G}_{\alpha\beta}&(\vec{k},k_{z}) =\\ \nonumber
   &\frac{k_{\alpha}k_{\beta}/k^{2}}{c_{11}(\vec{k},k_{z}) k^{2} + c_{44}(\vec{k},k_{z}) k_{z}^{2}} + \frac{\delta_{\alpha \beta} - k_{\alpha}k_{\beta}/k^{2}}{c_{66} k^{2} + c_{44}(\vec{k},k_{z}) k_{z}^{2}}
\end{align}
involves the compression ($c_{11}$), tilt ($c_{44}$), and shear ($c_{66}$) moduli. We neglect the first term on the right-hand side (since $c_{66} \ll c_{11}$) and use $c_{44}(\vec{k}, k_{z}) = c_{44}(\vec{0},0)/[1+\lambda^{2}(k^{2} + k_{z}^{2})]$. By further assuming $k^{2}\lambda^{2} \gg 1$ and $k^{2} \gg k_{z}^{2}$ the integration range over $k_{z}$ can be extended to infinity and the planar integration is limited to a circular Brillouin zone $k^{2} \leq 4\pi/a_{\sss \triangle}$. Within these approximations, we find
\begin{align}\label{eq:app-Cbar-of-a}
   \C(a_{z}) = \C_{0} \frac{\chi a_{z}/\ell_{h}}{\ln(1 + \chi a_{z}/\ell_{h})},
\end{align}
with $\chi$ a numerical of order unity. Analytical evaluation suggests  $\chi \approx \sqrt{\pi} \approx 1.77$.

We investigate the size-dependent elastic response by studying the linear regime with our TDGL simulations. More specifically, we determine the maximal deformation $u$ of the pinned vortex subject to a small Lorentz force $f_{L}$, see Fig.\ \ref{fig-InclVortLatt}. This deformation is evaluated as $u \equiv x(0) - x(L_{z}/2)$ with $x(z)$ being the vortex position at the height $z$ relative to the defect's center. Substituting the quantity $u$ into \eq \eqref{eq:el-force-balace} allows us to numerically evaluate the effective spring constant $\C$, see Fig.\ \ref{fig:Cbar}.

Simulations of an isotropic system show that the linear response of the vortex lattice to a small Lorentz force has noticeable dependence on the defect size, see Fig.\ \ref{fig:Cbar}(top). Using \eq \eqref{eq:app-Cbar-of-a}, we extract the point-size spring constant $\C_{0}$ (black symbols) and the numerical constant $\chi$ as fit parameters. For the anisotropic system with $\gamma = 5$ the extracted spring constant $\C(a_{z})$ shows a much stronger relative defect-size dependence, see Fig.\ \ref{fig:Cbar}(bottom). This suggests that even the smallest defect ($a = 2\xi$) may not be regarded as a pointlike object. Extracting $\C_{0}$ from fitting the numerical data with \eq \eqref{eq:app-Cbar-of-a}, further supports this observation. The dependences of $\C_{0}(B)$ for $\gamma = 1$ and $\gamma = 5$, agree well with the expected behavior. The solid curves in Fig.\ \ref{fig:Cbar} show  $1.15 \C_{0}(B)$, see \eq \eqref{eq:Cbar-of-0}, very close to the theoretical prediction. For both anisotropies,  $\chi$ lies in the range $2\text{ - }2.5$, close to the analytically-evaluated value.

For large defects $a\geq 4\xi$ the field dependence of $\C$ shows a pronounced downturn above the size-dependent magnetic field. The origin of this downturn is the proximity to the defect's double-occupancy [see Section \ref{sec:pin-breaking}]. An additional softening is expected when approaching $H_{c2}$. Accounting for all field dependences of the elastic constants,\cite{Brandt1986} it has been shown\cite{Willa2016} that $\C_{0}$ acquires an additional form factor $(1 - B/H_{c2})^{3/2}$.

\section{Campbell length}\label{sec:Campbell}
The $ac$ penetration depth of a low-amplitude, low-frequency field oscillation is another experimentally accessible quantity that allows to quantify vortex pinning. The pinned state responds to an external $ac$ perturbation $h_{ac}e^{-i \omega t}$ with an exponentially damped density modulation $h_{ac} e^{-i \omega t} e^{-X/\lambda_{\sss \mathrm{C}}}$. Here, $X$ measures macroscopic ($X \gg a_{\sss \triangle}$) distances away from the sample surface and the (Campbell) length $\lambda_{\sss \mathrm{C}}$ directly relates to the stiffness of the pinned vortex state; and hence on its microscopic origin.
\begin{figure}[htb]
\centering	
\includegraphics[width = .45\textwidth]{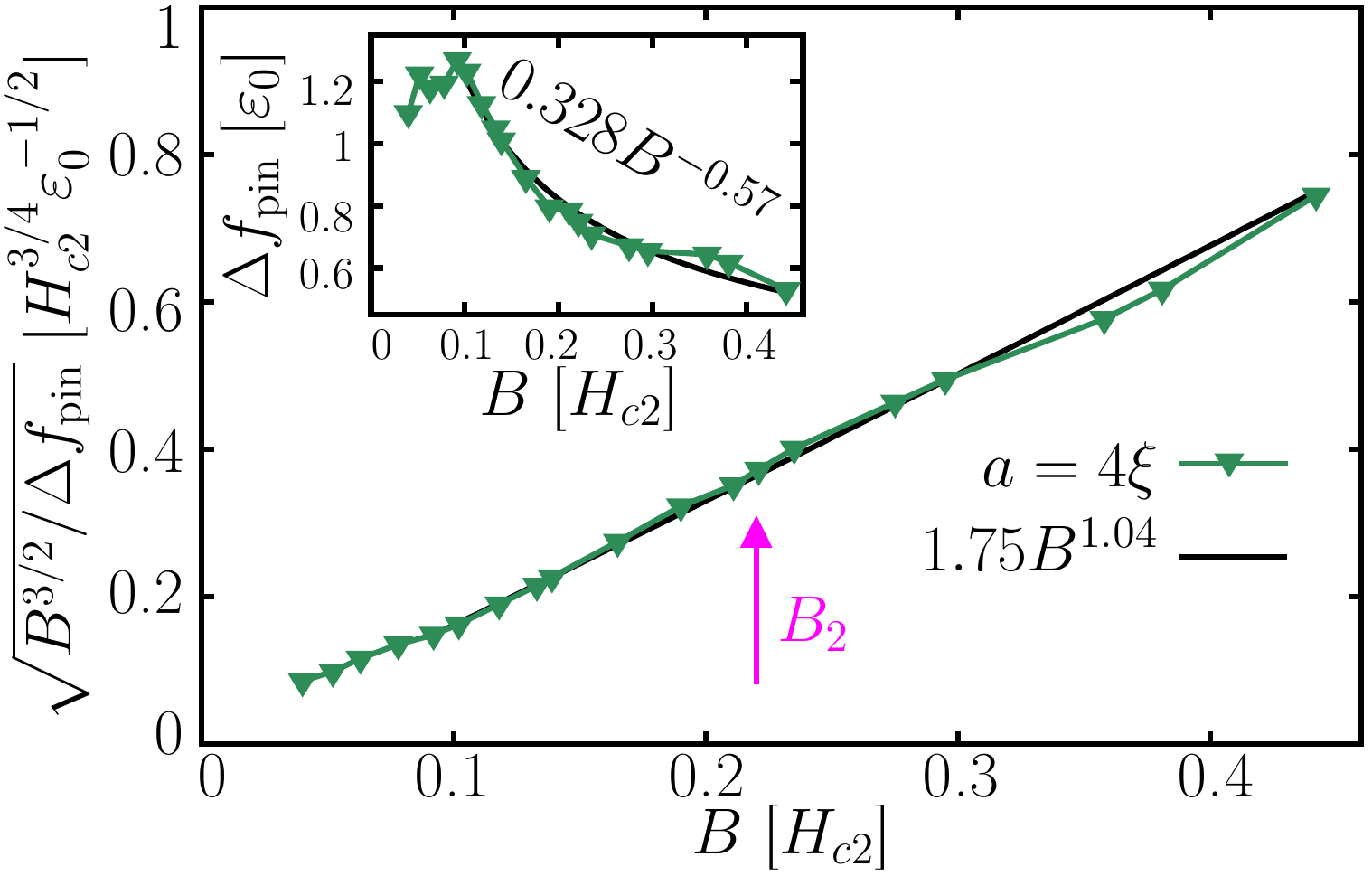}
\caption{At high fields, the quantity $ \sqrt{B^{3/2}/\Delta f_{\mathrm{pin}}}$ mimics the behavior of the Campbell length $\lambda_{\sss \mathrm{C}}$ for a system with a low density $n_{p}$ of defects, see \eq \eqref{eq:Lc}. Indeed, the proper average $\langle \partial f_{\mathrm{pin}} /\partial x \rangle$ measures\cite{Willa2016} the sum of the discontinuities (denoted by $\Delta f_{\mathrm{pin}}$) of the pinscape $f_{\mathrm{pin}}(x)$, see Fig.\ \ref{fig:fpin}(right). The field dependence $\lambda_{\sss \mathrm{C}}(B)$, shown here for $4\xi$, is almost linear. Another field dependence [$\lambda_{\sss \mathrm{C}}(B) \propto B^{1/2}$] is expected at low fields, where both the transverse length $2x_{-}$ (instead of $a_{\sss \triangle}$) and the force jump $\Delta f_{\mathrm{pin}}$ are weakly field-dependent.
}
\label{fig:lambdaC-of-B}
\end{figure}

Within the strong-pinning formalism, the Campbell length in the critical state has been shown to take the form \cite{Willa2015a, Willa2015b, Willa2016} $\lambda_{\sss \mathrm{C}}^{-2} \approx [4\pi n_{p} (2x_{-})/B \Phi_{0}] \langle \partial f_{\mathrm{pin}} /\partial x \rangle$. Whereas a single-valued force (weak pinning, $\kappa < 1$) yields $\langle \partial f_{\mathrm{pin}} /\partial x \rangle = 0$ due to symmetry, a finite average may only arise from discontinuities of the pinning force $f_{\mathrm{pin}}(x)$, i.e., under strong pinning conditions $\kappa > 1$. The above expression then simplifies to $\lambda_{\sss \mathrm{C}}^{-2} \approx [4\pi n_{p} (2x_{-})/B \Phi_{0}]\Delta f_{\mathrm{pin}}$, where $\Delta f_{\mathrm{pin}}$ measures the sum of the force discontinuities in the occupation of $f_{\mathrm{pin}}(x)$.

Our quasi-static simulations, see Section \ref{sec:dynamic}, probe the pinning profile as occupied in the critical [or Bean, or zero-field-cooled (zfc)] vortex state and hence allow to evaluate the corresponding Campbell length. The numerical evaluation of the force jumps suggest that $\Delta f_{\mathrm{pin}} \propto B^{-1/2}$ at high fields.
In the regime when the inclusion always occupied the transverse trapping distance $2x_{-}$ has to be replaced by $a_{\sss \triangle}$ and formula for $\lambda_{\sss \mathrm{C}}$ becomes
\begin{align}\label{eq:Lc}
   \lambda_{\sss \mathrm{C}}^{-2} &\approx \frac{4\pi n_{p} a_{\sss \triangle}}{B \Phi_{0}} \Delta f_{\mathrm{pin}}.
\end{align}
As a result the (zfc) Campbell length features an almost linear field dependence, see Fig.\ \ref{fig:lambdaC-of-B}. At the onset of double-occupancy, $B_{2} \approx 0.22 H_{c2}$, the single discontinuity in the force profile splits into two separate ones. Yet, the Campbell length (measuring the overall jump $\Delta f_{\mathrm{pin}}$) is remarkably insensitive to this transition.
The saturation of the force discontinuity $\Delta f_{\mathrm{pin}}$ at low fields point to the expected\cite{Willa2016} regime where the intervortex distance $a_{\sss \triangle}$ is larger than the transverse trapping length $2x_{-}$, and the (zfc) Campbell length grows as $\lambda_{\sss \mathrm{C}} \propto B^{1/2}$.

\vfill
\pagebreak

\bibliographystyle{apsrev4-1-titles}
\bibliography{SingleDefectPaper}

\begin{thebibliography}{60}%
\makeatletter
\providecommand \@ifxundefined [1]{%
 \@ifx{#1\undefined}
}%
\providecommand \@ifnum [1]{%
 \ifnum #1\expandafter \@firstoftwo
 \else \expandafter \@secondoftwo
 \fi
}%
\providecommand \@ifx [1]{%
 \ifx #1\expandafter \@firstoftwo
 \else \expandafter \@secondoftwo
 \fi
}%
\providecommand \natexlab [1]{#1}%
\providecommand \enquote  [1]{``#1''}%
\providecommand \bibnamefont  [1]{#1}%
\providecommand \bibfnamefont [1]{#1}%
\providecommand \citenamefont [1]{#1}%
\providecommand \href@noop [0]{\@secondoftwo}%
\providecommand \href [0]{\begingroup \@sanitize@url \@href}%
\providecommand \@href[1]{\@@startlink{#1}\@@href}%
\providecommand \@@href[1]{\endgroup#1\@@endlink}%
\providecommand \@sanitize@url [0]{\catcode `\\12\catcode `\$12\catcode
  `\&12\catcode `\#12\catcode `\^12\catcode `\_12\catcode `\%12\relax}%
\providecommand \@@startlink[1]{}%
\providecommand \@@endlink[0]{}%
\providecommand \url  [0]{\begingroup\@sanitize@url \@url }%
\providecommand \@url [1]{\endgroup\@href {#1}{\urlprefix }}%
\providecommand \urlprefix  [0]{URL }%
\providecommand \Eprint [0]{\href }%
\providecommand \doibase [0]{http://doi.org/}%
\providecommand \selectlanguage [0]{\@gobble}%
\providecommand \bibinfo  [0]{\@secondoftwo}%
\providecommand \bibfield  [0]{\@secondoftwo}%
\providecommand \translation [1]{[#1]}%
\providecommand \BibitemOpen [0]{}%
\providecommand \bibitemStop [0]{}%
\providecommand \bibitemNoStop [0]{.\EOS\space}%
\providecommand \EOS [0]{\spacefactor3000\relax}%
\providecommand \BibitemShut  [1]{\csname bibitem#1\endcsname}%
\let\auto@bib@innerbib\@empty
\bibitem [{\citenamefont {Abrikosov}(1957)}]{Abrikosov1957}%
  \BibitemOpen
  \bibfield  {author} {\bibinfo {author} {\bibfnamefont {A.~A.}\ \bibnamefont
  {Abrikosov}},\ }\bibfield  {title} {\emph {\bibinfo {title} {{On the magnetic
  properties of superconductors of the second group}}},\ }\href@noop {}
  {\bibfield  {journal} {\bibinfo  {journal} {{[Zh. Eksp. Teor. Fiz.
  \textbf{32}, 1442 (1957)]} JETP}\ }\textbf {\bibinfo {volume} {5}},\ \bibinfo
  {pages} {1174} (\bibinfo {year} {1957})}\BibitemShut {NoStop}%
\bibitem [{\citenamefont {Matsumoto}\ and\ \citenamefont
  {Mele}(2009)}]{Matsumoto2009}%
  \BibitemOpen
  \bibfield  {author} {\bibinfo {author} {\bibfnamefont {K.}~\bibnamefont
  {Matsumoto}}\ and\ \bibinfo {author} {\bibfnamefont {P.}~\bibnamefont
  {Mele}},\ }\bibfield  {title} {\emph {\bibinfo {title} {{Artificial pinning
  center technology to enhance vortex pinning in {YBCO} coated conductors}}},\
  }\href {\doibase 10.1088/0953-2048/23/1/014001} {\bibfield  {journal}
  {\bibinfo  {journal} {Supercond. Sci. Technol.}\ }\textbf {\bibinfo {volume}
  {23}},\ \bibinfo {pages} {014001} (\bibinfo {year} {2009})}\BibitemShut
  {NoStop}%
\bibitem [{\citenamefont {Obradors}\ and\ \citenamefont
  {Puig}(2014)}]{Obradors2014}%
  \BibitemOpen
  \bibfield  {author} {\bibinfo {author} {\bibfnamefont {X.}~\bibnamefont
  {Obradors}}\ and\ \bibinfo {author} {\bibfnamefont {T.}~\bibnamefont
  {Puig}},\ }\bibfield  {title} {\emph {\bibinfo {title} {{Coated conductors
  for power applications: materials challenges}}},\ }\href {\doibase
  10.1088/0953-2048/27/4/044003} {\bibfield  {journal} {\bibinfo  {journal}
  {Supercond. Sci. Technol.}\ }\textbf {\bibinfo {volume} {27}},\ \bibinfo
  {pages} {044003} (\bibinfo {year} {2014})}\BibitemShut {NoStop}%
\bibitem [{\citenamefont {Gurevich}(2014)}]{GurevichAnnRevCMP14}%
  \BibitemOpen
  \bibfield  {author} {\bibinfo {author} {\bibfnamefont {A.}~\bibnamefont
  {Gurevich}},\ }\bibfield  {title} {\emph {\bibinfo {title} {Challenges and
  opportunities for applications of unconventional superconductors}},\ }\href
  {\doibase 10.1146/annurev-conmatphys-031113-133822} {\bibfield  {journal}
  {\bibinfo  {journal} {Annu. Rev. Condens. Matter Phys.}\ }\textbf {\bibinfo
  {volume} {5}},\ \bibinfo {pages} {35} (\bibinfo {year} {2014})}\BibitemShut
  {NoStop}%
\bibitem [{\citenamefont {Kwok}\ \emph {et~al.}(2016)\citenamefont {Kwok},
  \citenamefont {Welp}, \citenamefont {Glatz}, \citenamefont {Koshelev},
  \citenamefont {Kihlstrom},\ and\ \citenamefont {Crabtree}}]{KwokRoPP2016}%
  \BibitemOpen
  \bibfield  {author} {\bibinfo {author} {\bibfnamefont {W.-K.}\ \bibnamefont
  {Kwok}}, \bibinfo {author} {\bibfnamefont {U.}~\bibnamefont {Welp}}, \bibinfo
  {author} {\bibfnamefont {A.}~\bibnamefont {Glatz}}, \bibinfo {author}
  {\bibfnamefont {A.~E.}\ \bibnamefont {Koshelev}}, \bibinfo {author}
  {\bibfnamefont {K.~J.}\ \bibnamefont {Kihlstrom}}, \ and\ \bibinfo {author}
  {\bibfnamefont {G.~W.}\ \bibnamefont {Crabtree}},\ }\bibfield  {title} {\emph
  {\bibinfo {title} {Vortices in high-performance high-temperature
  superconductors}},\ }\href {http://stacks.iop.org/0034-4885/79/i=11/a=116501}
  {\bibfield  {journal} {\bibinfo  {journal} {Rep. Prog. Phys.}\ }\textbf
  {\bibinfo {volume} {79}},\ \bibinfo {pages} {116501} (\bibinfo {year}
  {2016})}\BibitemShut {NoStop}%
\bibitem [{\citenamefont {Feighan}\ \emph {et~al.}(2017)\citenamefont
  {Feighan}, \citenamefont {Kursumovic},\ and\ \citenamefont
  {MacManus-Driscoll}}]{FeighanSUST17}%
  \BibitemOpen
  \bibfield  {author} {\bibinfo {author} {\bibfnamefont {J.~P.~F.}\
  \bibnamefont {Feighan}}, \bibinfo {author} {\bibfnamefont {A.}~\bibnamefont
  {Kursumovic}}, \ and\ \bibinfo {author} {\bibfnamefont {J.~L.}\ \bibnamefont
  {MacManus-Driscoll}},\ }\bibfield  {title} {\emph {\bibinfo {title}
  {Materials design for artificial pinning centres in superconductor {PLD}
  coated conductors}},\ }\href {\doibase 10.1088/1361-6668/aa90d1} {\bibfield
  {journal} {\bibinfo  {journal} {Supercond. Sci. Technol.}\ }\textbf {\bibinfo
  {volume} {30}},\ \bibinfo {pages} {123001} (\bibinfo {year}
  {2017})}\BibitemShut {NoStop}%
\bibitem [{\citenamefont {MacManus-Driscoll}\ \emph {et~al.}(2004)\citenamefont
  {MacManus-Driscoll}, \citenamefont {Foltyn}, \citenamefont {Jia},
  \citenamefont {Wang}, \citenamefont {Serquis}, \citenamefont {Maiorov},
  \citenamefont {Civale}, \citenamefont {Lin}, \citenamefont {Hawley},
  \citenamefont {Maley},\ and\ \citenamefont {Peterson}}]{MacManusAPL04}%
  \BibitemOpen
  \bibfield  {author} {\bibinfo {author} {\bibfnamefont {J.~L.}\ \bibnamefont
  {MacManus-Driscoll}}, \bibinfo {author} {\bibfnamefont {S.~R.}\ \bibnamefont
  {Foltyn}}, \bibinfo {author} {\bibfnamefont {Q.~X.}\ \bibnamefont {Jia}},
  \bibinfo {author} {\bibfnamefont {H.}~\bibnamefont {Wang}}, \bibinfo {author}
  {\bibfnamefont {A.}~\bibnamefont {Serquis}}, \bibinfo {author} {\bibfnamefont
  {B.}~\bibnamefont {Maiorov}}, \bibinfo {author} {\bibfnamefont
  {L.}~\bibnamefont {Civale}}, \bibinfo {author} {\bibfnamefont
  {Y.}~\bibnamefont {Lin}}, \bibinfo {author} {\bibfnamefont {M.~E.}\
  \bibnamefont {Hawley}}, \bibinfo {author} {\bibfnamefont {M.~P.}\
  \bibnamefont {Maley}}, \ and\ \bibinfo {author} {\bibfnamefont {D.~E.}\
  \bibnamefont {Peterson}},\ }\bibfield  {title} {\emph {\bibinfo {title}
  {Systematic enhancement of in-field critical current density with rare-earth
  ion size variance in superconducting rare-earth barium cuprate films}},\
  }\href {\doibase 10.1063/1.1766394} {\bibfield  {journal} {\bibinfo
  {journal} {Appl. Phys. Lett.}\ }\textbf {\bibinfo {volume} {84}},\ \bibinfo
  {pages} {5329} (\bibinfo {year} {2004})}\BibitemShut {NoStop}%
\bibitem [{\citenamefont {Haugan}\ \emph {et~al.}(2004)\citenamefont {Haugan},
  \citenamefont {Barnes}, \citenamefont {Wheeler}, \citenamefont
  {Meisenkothen},\ and\ \citenamefont {Sumption}}]{HauganNat04}%
  \BibitemOpen
  \bibfield  {author} {\bibinfo {author} {\bibfnamefont {T.}~\bibnamefont
  {Haugan}}, \bibinfo {author} {\bibfnamefont {P.~N.}\ \bibnamefont {Barnes}},
  \bibinfo {author} {\bibfnamefont {R.}~\bibnamefont {Wheeler}}, \bibinfo
  {author} {\bibfnamefont {F.}~\bibnamefont {Meisenkothen}}, \ and\ \bibinfo
  {author} {\bibfnamefont {M.}~\bibnamefont {Sumption}},\ }\bibfield  {title}
  {\emph {\bibinfo {title} {Addition of nanoparticle dispersions to enhance
  flux pinning of the {$\mathrm{Y}\mathrm{Ba}_2\mathrm{Cu}_3\mathrm{O}_{7-x}$}
  superconductor}},\ }\href {\doibase 10.1038/nature02792} {\bibfield
  {journal} {\bibinfo  {journal} {Nature}\ }\textbf {\bibinfo {volume} {430}},\
  \bibinfo {pages} {867} (\bibinfo {year} {2004})}\BibitemShut {NoStop}%
\bibitem [{\citenamefont {Song}\ \emph {et~al.}(2006)\citenamefont {Song},
  \citenamefont {Chen}, \citenamefont {Kim}, \citenamefont {Feldmann},
  \citenamefont {Larbalestier}, \citenamefont {Reeves}, \citenamefont {Xie},\
  and\ \citenamefont {Selvamanickam}}]{SongAPL06}%
  \BibitemOpen
  \bibfield  {author} {\bibinfo {author} {\bibfnamefont {X.}~\bibnamefont
  {Song}}, \bibinfo {author} {\bibfnamefont {Z.}~\bibnamefont {Chen}}, \bibinfo
  {author} {\bibfnamefont {S.-I.}\ \bibnamefont {Kim}}, \bibinfo {author}
  {\bibfnamefont {D.~M.}\ \bibnamefont {Feldmann}}, \bibinfo {author}
  {\bibfnamefont {D.}~\bibnamefont {Larbalestier}}, \bibinfo {author}
  {\bibfnamefont {J.}~\bibnamefont {Reeves}}, \bibinfo {author} {\bibfnamefont
  {Y.}~\bibnamefont {Xie}}, \ and\ \bibinfo {author} {\bibfnamefont
  {V.}~\bibnamefont {Selvamanickam}},\ }\bibfield  {title} {\emph {\bibinfo
  {title} {Evidence for strong flux pinning by small, dense nanoprecipitates in
  a {$\mathrm{Sm}$}-doped
  {$\mathrm{Y}\mathrm{Ba}_2\mathrm{Cu}_3\mathrm{O}_{7-\delta}$} coated
  conductor}},\ }\href {\doibase 10.1063/1.2206989} {\bibfield  {journal}
  {\bibinfo  {journal} {Appl. Phys. Lett.}\ }\textbf {\bibinfo {volume} {88}},\
  \bibinfo {pages} {212508} (\bibinfo {year} {2006})}\BibitemShut {NoStop}%
\bibitem [{\citenamefont {Gutierrez}\ \emph {et~al.}(2007)\citenamefont
  {Gutierrez}, \citenamefont {Llordes}, \citenamefont {Gazquez}, \citenamefont
  {Gibert}, \citenamefont {Roma}, \citenamefont {Pomar}, \citenamefont
  {Sandiumenge}, \citenamefont {Mestres}, \citenamefont {Puig},\ and\
  \citenamefont {Obradors}}]{GutierrezNatMat07}%
  \BibitemOpen
  \bibfield  {author} {\bibinfo {author} {\bibfnamefont {J.}~\bibnamefont
  {Gutierrez}}, \bibinfo {author} {\bibfnamefont {A.}~\bibnamefont {Llordes}},
  \bibinfo {author} {\bibfnamefont {J.}~\bibnamefont {Gazquez}}, \bibinfo
  {author} {\bibfnamefont {M.}~\bibnamefont {Gibert}}, \bibinfo {author}
  {\bibfnamefont {N.}~\bibnamefont {Roma}}, \bibinfo {author} {\bibfnamefont
  {A.}~\bibnamefont {Pomar}}, \bibinfo {author} {\bibfnamefont
  {F.}~\bibnamefont {Sandiumenge}}, \bibinfo {author} {\bibfnamefont
  {N.}~\bibnamefont {Mestres}}, \bibinfo {author} {\bibfnamefont
  {T.}~\bibnamefont {Puig}}, \ and\ \bibinfo {author} {\bibfnamefont
  {X.}~\bibnamefont {Obradors}},\ }\bibfield  {title} {\emph {\bibinfo {title}
  {Strong isotropic flux pinning in solution-derived
  {$\mathrm{Y}\mathrm{Ba}_2\mathrm{Cu}_3\mathrm{O}_{7-x}$} nanocomposite
  superconductor films}},\ }\href {\doibase 10.1038/nmat1893} {\bibfield
  {journal} {\bibinfo  {journal} {Nat. Mater.}\ }\textbf {\bibinfo {volume}
  {6}},\ \bibinfo {pages} {367} (\bibinfo {year} {2007})}\BibitemShut {NoStop}%
\bibitem [{\citenamefont {Yamasaki}\ \emph {et~al.}(2008)\citenamefont
  {Yamasaki}, \citenamefont {Ohki}, \citenamefont {Yamada}, \citenamefont
  {Nakagawa},\ and\ \citenamefont {Mawatari}}]{YamasakiSUST08}%
  \BibitemOpen
  \bibfield  {author} {\bibinfo {author} {\bibfnamefont {H.}~\bibnamefont
  {Yamasaki}}, \bibinfo {author} {\bibfnamefont {K.}~\bibnamefont {Ohki}},
  \bibinfo {author} {\bibfnamefont {H.}~\bibnamefont {Yamada}}, \bibinfo
  {author} {\bibfnamefont {Y.}~\bibnamefont {Nakagawa}}, \ and\ \bibinfo
  {author} {\bibfnamefont {Y.}~\bibnamefont {Mawatari}},\ }\bibfield  {title}
  {\emph {\bibinfo {title} {Strong flux pinning in
  {$\mathrm{Y}\mathrm{Ba}_2\mathrm{Cu}_3\mathrm{O}_7$} thin films due to
  nanometer-sized precipitates}},\ }\href {\doibase
  10.1088/0953-2048/21/12/125011} {\bibfield  {journal} {\bibinfo  {journal}
  {Supercond. Sci. Technol.}\ }\textbf {\bibinfo {volume} {21}},\ \bibinfo
  {pages} {125011} (\bibinfo {year} {2008})}\BibitemShut {NoStop}%
\bibitem [{\citenamefont {Yamasaki}(2016)}]{YamasakiSUST16}%
  \BibitemOpen
  \bibfield  {author} {\bibinfo {author} {\bibfnamefont {H.}~\bibnamefont
  {Yamasaki}},\ }\bibfield  {title} {\emph {\bibinfo {title} {Effect of
  particle size on the flux pinning properties of
  {$\mathrm{Y}\mathrm{Ba}_2\mathrm{Cu}_3\mathrm{O}_{7-\delta}$} thin films
  containing fine {$\mathrm{Y}_2\mathrm{O}_3$} nanoprecipitates}},\ }\href
  {\doibase 10.1088/0953-2048/29/6/065005} {\bibfield  {journal} {\bibinfo
  {journal} {Supercond. Sci. Technol.}\ }\textbf {\bibinfo {volume} {29}},\
  \bibinfo {pages} {065005} (\bibinfo {year} {2016})}\BibitemShut {NoStop}%
\bibitem [{\citenamefont {Polat}\ \emph {et~al.}(2011)\citenamefont {Polat},
  \citenamefont {Sinclair}, \citenamefont {Zuev}, \citenamefont {Thompson},
  \citenamefont {Christen}, \citenamefont {Cook}, \citenamefont {Kumar},
  \citenamefont {Chen},\ and\ \citenamefont {Selvamanickam}}]{PolatPhysRevB11}%
  \BibitemOpen
  \bibfield  {author} {\bibinfo {author} {\bibfnamefont {O.}~\bibnamefont
  {Polat}}, \bibinfo {author} {\bibfnamefont {J.~W.}\ \bibnamefont {Sinclair}},
  \bibinfo {author} {\bibfnamefont {Y.~L.}\ \bibnamefont {Zuev}}, \bibinfo
  {author} {\bibfnamefont {J.~R.}\ \bibnamefont {Thompson}}, \bibinfo {author}
  {\bibfnamefont {D.~K.}\ \bibnamefont {Christen}}, \bibinfo {author}
  {\bibfnamefont {S.~W.}\ \bibnamefont {Cook}}, \bibinfo {author}
  {\bibfnamefont {D.}~\bibnamefont {Kumar}}, \bibinfo {author} {\bibfnamefont
  {Y.}~\bibnamefont {Chen}}, \ and\ \bibinfo {author} {\bibfnamefont
  {V.}~\bibnamefont {Selvamanickam}},\ }\bibfield  {title} {\emph {\bibinfo
  {title} {Thickness dependence of magnetic relaxation and {$E$-$J$}
  characteristics in superconducting
  {$\mathrm{(Gd\text{-}Y)\text{-}Ba\text{-}Cu\text{-}O}$} films with strong
  vortex pinning}},\ }\href {\doibase 10.1103/PhysRevB.84.024519} {\bibfield
  {journal} {\bibinfo  {journal} {Phys. Rev. B}\ }\textbf {\bibinfo {volume}
  {84}},\ \bibinfo {pages} {024519} (\bibinfo {year} {2011})}\BibitemShut
  {NoStop}%
\bibitem [{\citenamefont {Miura}\ \emph {et~al.}(2011)\citenamefont {Miura},
  \citenamefont {Maiorov}, \citenamefont {Baily}, \citenamefont {Haberkorn},
  \citenamefont {Willis}, \citenamefont {Marken}, \citenamefont {Izumi},
  \citenamefont {Shiohara},\ and\ \citenamefont {Civale}}]{MiuraPhysRevB11}%
  \BibitemOpen
  \bibfield  {author} {\bibinfo {author} {\bibfnamefont {M.}~\bibnamefont
  {Miura}}, \bibinfo {author} {\bibfnamefont {B.}~\bibnamefont {Maiorov}},
  \bibinfo {author} {\bibfnamefont {S.~A.}\ \bibnamefont {Baily}}, \bibinfo
  {author} {\bibfnamefont {N.}~\bibnamefont {Haberkorn}}, \bibinfo {author}
  {\bibfnamefont {J.~O.}\ \bibnamefont {Willis}}, \bibinfo {author}
  {\bibfnamefont {K.}~\bibnamefont {Marken}}, \bibinfo {author} {\bibfnamefont
  {T.}~\bibnamefont {Izumi}}, \bibinfo {author} {\bibfnamefont
  {Y.}~\bibnamefont {Shiohara}}, \ and\ \bibinfo {author} {\bibfnamefont
  {L.}~\bibnamefont {Civale}},\ }\bibfield  {title} {\emph {\bibinfo {title}
  {Mixed pinning landscape in nanoparticle-introduced
  {$\mathrm{Y}\mathrm{Gd}\mathrm{Ba}_2\mathrm{Cu}_3\mathrm{O}_{y}$} films grown
  by metal organic deposition}},\ }\href {\doibase 10.1103/PhysRevB.83.184519}
  {\bibfield  {journal} {\bibinfo  {journal} {Phys. Rev. B}\ }\textbf {\bibinfo
  {volume} {83}},\ \bibinfo {pages} {184519} (\bibinfo {year}
  {2011})}\BibitemShut {NoStop}%
\bibitem [{\citenamefont {Miura}\ \emph {et~al.}(2013)\citenamefont {Miura},
  \citenamefont {Maiorov}, \citenamefont {Willis}, \citenamefont {Kato},
  \citenamefont {Sato}, \citenamefont {Izumi}, \citenamefont {Shiohara},\ and\
  \citenamefont {Civale}}]{Miura2013}%
  \BibitemOpen
  \bibfield  {author} {\bibinfo {author} {\bibfnamefont {M.}~\bibnamefont
  {Miura}}, \bibinfo {author} {\bibfnamefont {B.}~\bibnamefont {Maiorov}},
  \bibinfo {author} {\bibfnamefont {J.~O.}\ \bibnamefont {Willis}}, \bibinfo
  {author} {\bibfnamefont {T.}~\bibnamefont {Kato}}, \bibinfo {author}
  {\bibfnamefont {M.}~\bibnamefont {Sato}}, \bibinfo {author} {\bibfnamefont
  {T.}~\bibnamefont {Izumi}}, \bibinfo {author} {\bibfnamefont
  {Y.}~\bibnamefont {Shiohara}}, \ and\ \bibinfo {author} {\bibfnamefont
  {L.}~\bibnamefont {Civale}},\ }\bibfield  {title} {\emph {\bibinfo {title}
  {{The effects of density and size of {$\mathrm{Ba}M\mathrm{O}_3$ ($M=
  \mathrm{Zr}, \mathrm{Nb}, \mathrm{Sn}$)} nanoparticles on the vortex glassy
  and liquid phase in
  {$(\mathrm{Y},\mathrm{Gd})\mathrm{Ba}_2\mathrm{Cu}_3\mathrm{O}_y$} coated
  conductors}}},\ }\href {\doibase 10.1088/0953-2048/26/3/035008} {\bibfield
  {journal} {\bibinfo  {journal} {Supercond. Sci. Technol.}\ }\textbf {\bibinfo
  {volume} {26}},\ \bibinfo {pages} {035008} (\bibinfo {year}
  {2013})}\BibitemShut {NoStop}%
\bibitem [{\citenamefont {Miura}\ \emph {et~al.}(2017)\citenamefont {Miura},
  \citenamefont {Maiorov}, \citenamefont {Sato}, \citenamefont {Kanai},
  \citenamefont {Kato}, \citenamefont {Kato}, \citenamefont {Izumi},
  \citenamefont {Awaji}, \citenamefont {Mele}, \citenamefont {Kiuchi},\ and\
  \citenamefont {Matsushita}}]{MiuraNPGAsMat17}%
  \BibitemOpen
  \bibfield  {author} {\bibinfo {author} {\bibfnamefont {M.}~\bibnamefont
  {Miura}}, \bibinfo {author} {\bibfnamefont {B.}~\bibnamefont {Maiorov}},
  \bibinfo {author} {\bibfnamefont {M.}~\bibnamefont {Sato}}, \bibinfo {author}
  {\bibfnamefont {M.}~\bibnamefont {Kanai}}, \bibinfo {author} {\bibfnamefont
  {T.}~\bibnamefont {Kato}}, \bibinfo {author} {\bibfnamefont {T.}~\bibnamefont
  {Kato}}, \bibinfo {author} {\bibfnamefont {T.}~\bibnamefont {Izumi}},
  \bibinfo {author} {\bibfnamefont {S.}~\bibnamefont {Awaji}}, \bibinfo
  {author} {\bibfnamefont {P.}~\bibnamefont {Mele}}, \bibinfo {author}
  {\bibfnamefont {M.}~\bibnamefont {Kiuchi}}, \ and\ \bibinfo {author}
  {\bibfnamefont {T.}~\bibnamefont {Matsushita}},\ }\bibfield  {title} {\emph
  {\bibinfo {title} {Tuning nanoparticle size for enhanced functionality in
  perovskite thin films deposited by metal organic deposition}},\ }\href
  {\doibase 10.1038/am.2017.197} {\bibfield  {journal} {\bibinfo  {journal}
  {NPG Asia Mater.}\ }\textbf {\bibinfo {volume} {9}},\ \bibinfo {pages} {e447}
  (\bibinfo {year} {2017})}\BibitemShut {NoStop}%
\bibitem [{\citenamefont {Goyal}\ \emph {et~al.}(2005)\citenamefont {Goyal},
  \citenamefont {Kang}, \citenamefont {Leonard}, \citenamefont {Martin},
  \citenamefont {Gapud}, \citenamefont {Varela}, \citenamefont {Paranthaman},
  \citenamefont {Ijaduola}, \citenamefont {Specht}, \citenamefont {Thompson},
  \citenamefont {Christen}, \citenamefont {Pennycook},\ and\ \citenamefont
  {List}}]{GoyalSUST05}%
  \BibitemOpen
  \bibfield  {author} {\bibinfo {author} {\bibfnamefont {A.}~\bibnamefont
  {Goyal}}, \bibinfo {author} {\bibfnamefont {S.}~\bibnamefont {Kang}},
  \bibinfo {author} {\bibfnamefont {K.~J.}\ \bibnamefont {Leonard}}, \bibinfo
  {author} {\bibfnamefont {P.~M.}\ \bibnamefont {Martin}}, \bibinfo {author}
  {\bibfnamefont {A.~A.}\ \bibnamefont {Gapud}}, \bibinfo {author}
  {\bibfnamefont {M.}~\bibnamefont {Varela}}, \bibinfo {author} {\bibfnamefont
  {M.}~\bibnamefont {Paranthaman}}, \bibinfo {author} {\bibfnamefont {A.~O.}\
  \bibnamefont {Ijaduola}}, \bibinfo {author} {\bibfnamefont {E.~D.}\
  \bibnamefont {Specht}}, \bibinfo {author} {\bibfnamefont {J.~R.}\
  \bibnamefont {Thompson}}, \bibinfo {author} {\bibfnamefont {D.~K.}\
  \bibnamefont {Christen}}, \bibinfo {author} {\bibfnamefont {S.~J.}\
  \bibnamefont {Pennycook}}, \ and\ \bibinfo {author} {\bibfnamefont {F.~A.}\
  \bibnamefont {List}},\ }\bibfield  {title} {\emph {\bibinfo {title}
  {Irradiation-free, columnar defects comprised of self-assembled nanodots and
  nanorods resulting in strongly enhanced flux-pinning in
  {$\mathrm{Y}\mathrm{Ba}_2\mathrm{Cu}_3\mathrm{O}_{7-\delta}$} films}},\
  }\href {\doibase 10.1088/0953-2048/18/11/021} {\bibfield  {journal} {\bibinfo
   {journal} {Supercond. Sci. Technol.}\ }\textbf {\bibinfo {volume} {18}},\
  \bibinfo {pages} {1533} (\bibinfo {year} {2005})}\BibitemShut {NoStop}%
\bibitem [{\citenamefont {Kang}\ \emph {et~al.}(2006)\citenamefont {Kang},
  \citenamefont {Goyal}, \citenamefont {Li}, \citenamefont {Gapud},
  \citenamefont {Martin}, \citenamefont {Heatherly}, \citenamefont {Thompson},
  \citenamefont {Christen}, \citenamefont {List}, \citenamefont {Paranthaman},\
  and\ \citenamefont {Lee}}]{KangSci06}%
  \BibitemOpen
  \bibfield  {author} {\bibinfo {author} {\bibfnamefont {S.}~\bibnamefont
  {Kang}}, \bibinfo {author} {\bibfnamefont {A.}~\bibnamefont {Goyal}},
  \bibinfo {author} {\bibfnamefont {J.}~\bibnamefont {Li}}, \bibinfo {author}
  {\bibfnamefont {A.~A.}\ \bibnamefont {Gapud}}, \bibinfo {author}
  {\bibfnamefont {P.~M.}\ \bibnamefont {Martin}}, \bibinfo {author}
  {\bibfnamefont {L.}~\bibnamefont {Heatherly}}, \bibinfo {author}
  {\bibfnamefont {J.~R.}\ \bibnamefont {Thompson}}, \bibinfo {author}
  {\bibfnamefont {D.~K.}\ \bibnamefont {Christen}}, \bibinfo {author}
  {\bibfnamefont {F.~A.}\ \bibnamefont {List}}, \bibinfo {author}
  {\bibfnamefont {M.}~\bibnamefont {Paranthaman}}, \ and\ \bibinfo {author}
  {\bibfnamefont {D.~F.}\ \bibnamefont {Lee}},\ }\bibfield  {title} {\emph
  {\bibinfo {title} {High-performance high-{$T_c$} superconducting wires}},\
  }\href {\doibase 10.1126/science.1124872} {\bibfield  {journal} {\bibinfo
  {journal} {Science}\ }\textbf {\bibinfo {volume} {311}},\ \bibinfo {pages}
  {1911} (\bibinfo {year} {2006})}\BibitemShut {NoStop}%
\bibitem [{\citenamefont {Mele}\ \emph {et~al.}(2008)\citenamefont {Mele},
  \citenamefont {Matsumoto}, \citenamefont {Horide}, \citenamefont {Ichinose},
  \citenamefont {Mukaida}, \citenamefont {Yoshida}, \citenamefont {Horii},\
  and\ \citenamefont {Kita}}]{Mele2008}%
  \BibitemOpen
  \bibfield  {author} {\bibinfo {author} {\bibfnamefont {P.}~\bibnamefont
  {Mele}}, \bibinfo {author} {\bibfnamefont {K.}~\bibnamefont {Matsumoto}},
  \bibinfo {author} {\bibfnamefont {T.}~\bibnamefont {Horide}}, \bibinfo
  {author} {\bibfnamefont {A.}~\bibnamefont {Ichinose}}, \bibinfo {author}
  {\bibfnamefont {M.}~\bibnamefont {Mukaida}}, \bibinfo {author} {\bibfnamefont
  {Y.}~\bibnamefont {Yoshida}}, \bibinfo {author} {\bibfnamefont
  {S.}~\bibnamefont {Horii}}, \ and\ \bibinfo {author} {\bibfnamefont
  {R.}~\bibnamefont {Kita}},\ }\bibfield  {title} {\emph {\bibinfo {title}
  {Ultra-high flux pinning properties of {$\mathrm{Ba}M\mathrm{O}_3$}-doped
  {$\mathrm{Y}\mathrm{Ba}_2\mathrm{Cu}_3\mathrm{O}_{7-x}$} thin films {$(M =
  \mathrm{Zr}, \mathrm{Sn})$}}},\ }\href {\doibase
  10.1088/0953-2048/21/3/032002} {\bibfield  {journal} {\bibinfo  {journal}
  {Supercond. Sci. Technol.}\ }\textbf {\bibinfo {volume} {21}},\ \bibinfo
  {pages} {032002} (\bibinfo {year} {2008})}\BibitemShut {NoStop}%
\bibitem [{\citenamefont {Selvamanickam}\ \emph {et~al.}(2015)\citenamefont
  {Selvamanickam}, \citenamefont {Gharahcheshmeh}, \citenamefont {Xu},
  \citenamefont {Galstyan}, \citenamefont {Delgado},\ and\ \citenamefont
  {Cantoni}}]{Selva2015b}%
  \BibitemOpen
  \bibfield  {author} {\bibinfo {author} {\bibfnamefont {V.}~\bibnamefont
  {Selvamanickam}}, \bibinfo {author} {\bibfnamefont {M.~H.}\ \bibnamefont
  {Gharahcheshmeh}}, \bibinfo {author} {\bibfnamefont {a.}~\bibnamefont {Xu}},
  \bibinfo {author} {\bibfnamefont {E.}~\bibnamefont {Galstyan}}, \bibinfo
  {author} {\bibfnamefont {L.}~\bibnamefont {Delgado}}, \ and\ \bibinfo
  {author} {\bibfnamefont {C.}~\bibnamefont {Cantoni}},\ }\bibfield  {title}
  {\emph {\bibinfo {title} {High critical currents in heavily doped
  {$(\mathrm{Gd},\mathrm{Y})\mathrm{Ba}_2\mathrm{Cu}_3\mathrm{O}_x$}
  superconductor tapes}},\ }\href {\doibase 10.1063/1.4906205} {\bibfield
  {journal} {\bibinfo  {journal} {Appl. Phys. Lett.}\ }\textbf {\bibinfo
  {volume} {106}},\ \bibinfo {pages} {032601} (\bibinfo {year}
  {2015})}\BibitemShut {NoStop}%
\bibitem [{\citenamefont {Awaji}\ \emph {et~al.}(2017)\citenamefont {Awaji},
  \citenamefont {Tsuchiya}, \citenamefont {Miura}, \citenamefont {Ichino},
  \citenamefont {Yoshida},\ and\ \citenamefont {Matsumoto}}]{AwajiSuST2017}%
  \BibitemOpen
  \bibfield  {author} {\bibinfo {author} {\bibfnamefont {S.}~\bibnamefont
  {Awaji}}, \bibinfo {author} {\bibfnamefont {Y.}~\bibnamefont {Tsuchiya}},
  \bibinfo {author} {\bibfnamefont {S.}~\bibnamefont {Miura}}, \bibinfo
  {author} {\bibfnamefont {Y.}~\bibnamefont {Ichino}}, \bibinfo {author}
  {\bibfnamefont {Y.}~\bibnamefont {Yoshida}}, \ and\ \bibinfo {author}
  {\bibfnamefont {K.}~\bibnamefont {Matsumoto}},\ }\bibfield  {title} {\emph
  {\bibinfo {title} {$c$-axis correlated pinning mechanism in vortex liquid and
  solid phases for {$\mathrm{Sm}$123} film with well-aligned
  {$\mathrm{Ba}\mathrm{Hf}\mathrm{O}_3$} nanorods}},\ }\href {\doibase
  10.1088/1361-6668/aa8d10} {\bibfield  {journal} {\bibinfo  {journal}
  {Supercond. Sci. Technol.}\ }\textbf {\bibinfo {volume} {30}},\ \bibinfo
  {pages} {114005} (\bibinfo {year} {2017})}\BibitemShut {NoStop}%
\bibitem [{\citenamefont {Jia}\ \emph {et~al.}(2013)\citenamefont {Jia},
  \citenamefont {LeRoux}, \citenamefont {Miller}, \citenamefont {Wen},
  \citenamefont {Kwok}, \citenamefont {Welp}, \citenamefont {Rupich},
  \citenamefont {Li}, \citenamefont {Sathyamurthy}, \citenamefont {Fleshler},
  \citenamefont {Malozemoff}, \citenamefont {Kayani}, \citenamefont
  {Ayala-Valenzuela},\ and\ \citenamefont {Civale}}]{JiaAPL13}%
  \BibitemOpen
  \bibfield  {author} {\bibinfo {author} {\bibfnamefont {Y.}~\bibnamefont
  {Jia}}, \bibinfo {author} {\bibfnamefont {M.}~\bibnamefont {LeRoux}},
  \bibinfo {author} {\bibfnamefont {D.~J.}\ \bibnamefont {Miller}}, \bibinfo
  {author} {\bibfnamefont {J.~G.}\ \bibnamefont {Wen}}, \bibinfo {author}
  {\bibfnamefont {W.~K.}\ \bibnamefont {Kwok}}, \bibinfo {author}
  {\bibfnamefont {U.}~\bibnamefont {Welp}}, \bibinfo {author} {\bibfnamefont
  {M.~W.}\ \bibnamefont {Rupich}}, \bibinfo {author} {\bibfnamefont
  {X.}~\bibnamefont {Li}}, \bibinfo {author} {\bibfnamefont {S.}~\bibnamefont
  {Sathyamurthy}}, \bibinfo {author} {\bibfnamefont {S.}~\bibnamefont
  {Fleshler}}, \bibinfo {author} {\bibfnamefont {A.~P.}\ \bibnamefont
  {Malozemoff}}, \bibinfo {author} {\bibfnamefont {A.}~\bibnamefont {Kayani}},
  \bibinfo {author} {\bibfnamefont {O.}~\bibnamefont {Ayala-Valenzuela}}, \
  and\ \bibinfo {author} {\bibfnamefont {L.}~\bibnamefont {Civale}},\
  }\bibfield  {title} {\emph {\bibinfo {title} {Doubling the critical current
  density of high temperature superconducting coated conductors through proton
  irradiation}},\ }\href {\doibase 10.1063/1.4821440} {\bibfield  {journal}
  {\bibinfo  {journal} {Appl. Phys. Lett.}\ }\textbf {\bibinfo {volume}
  {103}},\ \bibinfo {eid} {122601} (\bibinfo {year} {2013})}\BibitemShut
  {NoStop}%
\bibitem [{\citenamefont {Kihlstrom}\ \emph {et~al.}(2013)\citenamefont
  {Kihlstrom}, \citenamefont {Fang}, \citenamefont {Jia}, \citenamefont {Shen},
  \citenamefont {Koshelev}, \citenamefont {Welp}, \citenamefont {Crabtree},
  \citenamefont {Kwok}, \citenamefont {Kayani}, \citenamefont {Zhu},\ and\
  \citenamefont {Wen}}]{Kihlstrom2013}%
  \BibitemOpen
  \bibfield  {author} {\bibinfo {author} {\bibfnamefont {K.~J.}\ \bibnamefont
  {Kihlstrom}}, \bibinfo {author} {\bibfnamefont {L.}~\bibnamefont {Fang}},
  \bibinfo {author} {\bibfnamefont {Y.}~\bibnamefont {Jia}}, \bibinfo {author}
  {\bibfnamefont {B.}~\bibnamefont {Shen}}, \bibinfo {author} {\bibfnamefont
  {A.~E.}\ \bibnamefont {Koshelev}}, \bibinfo {author} {\bibfnamefont
  {U.}~\bibnamefont {Welp}}, \bibinfo {author} {\bibfnamefont {G.~W.}\
  \bibnamefont {Crabtree}}, \bibinfo {author} {\bibfnamefont {W.-K.}\
  \bibnamefont {Kwok}}, \bibinfo {author} {\bibfnamefont {A.}~\bibnamefont
  {Kayani}}, \bibinfo {author} {\bibfnamefont {S.~F.}\ \bibnamefont {Zhu}}, \
  and\ \bibinfo {author} {\bibfnamefont {H.-H.}\ \bibnamefont {Wen}},\
  }\bibfield  {title} {\emph {\bibinfo {title} {High-field critical current
  enhancement by irradiation induced correlated and random defects in
  {$(\mathrm{Ba}_{0.6}\mathrm{K}_{0.4})\mathrm{Fe}_2\mathrm{As}_2$}}},\ }\href
  {\doibase 10.1063/1.4829524} {\bibfield  {journal} {\bibinfo  {journal}
  {Appl. Phys. Lett.}\ }\textbf {\bibinfo {volume} {103}},\ \bibinfo {pages}
  {202601} (\bibinfo {year} {2013})}\BibitemShut {NoStop}%
\bibitem [{\citenamefont {Haberkorn}\ \emph {et~al.}(2015)\citenamefont
  {Haberkorn}, \citenamefont {Kim}, \citenamefont {Su\'{a}rez}, \citenamefont
  {Lee},\ and\ \citenamefont {Moon}}]{Haberkorn2015}%
  \BibitemOpen
  \bibfield  {author} {\bibinfo {author} {\bibfnamefont {N.}~\bibnamefont
  {Haberkorn}}, \bibinfo {author} {\bibfnamefont {J.}~\bibnamefont {Kim}},
  \bibinfo {author} {\bibfnamefont {S.}~\bibnamefont {Su\'{a}rez}}, \bibinfo
  {author} {\bibfnamefont {J.-H.}\ \bibnamefont {Lee}}, \ and\ \bibinfo
  {author} {\bibfnamefont {S.~H.}\ \bibnamefont {Moon}},\ }\bibfield  {title}
  {\emph {\bibinfo {title} {{Influence of random point defects introduced by
  proton irradiation on the flux creep rates and magnetic field dependence of
  the critical current density {$J_\mathrm{c}$} of co-evaporated
  {$\mathrm{Gd}\mathrm{Ba}_{2}\mathrm{Cu}_{3}\mathrm{O}_{7-\delta}$} coated
  conductors}}},\ }\href {\doibase 10.1088/0953-2048/28/12/125007} {\bibfield
  {journal} {\bibinfo  {journal} {Supercond. Sci. Technol.}\ }\textbf {\bibinfo
  {volume} {28}},\ \bibinfo {pages} {125007} (\bibinfo {year}
  {2015})}\BibitemShut {NoStop}%
\bibitem [{\citenamefont {Leroux}\ \emph {et~al.}(2015)\citenamefont {Leroux},
  \citenamefont {Kihlstrom}, \citenamefont {Holleis}, \citenamefont {Rupich},
  \citenamefont {Sathyamurthy}, \citenamefont {Fleshler}, \citenamefont
  {Sheng}, \citenamefont {Miller}, \citenamefont {Eley}, \citenamefont
  {Civale}, \citenamefont {Kayani}, \citenamefont {Niraula}, \citenamefont
  {Welp},\ and\ \citenamefont {Kwok}}]{Leroux2015}%
  \BibitemOpen
  \bibfield  {author} {\bibinfo {author} {\bibfnamefont {M.}~\bibnamefont
  {Leroux}}, \bibinfo {author} {\bibfnamefont {K.~J.}\ \bibnamefont
  {Kihlstrom}}, \bibinfo {author} {\bibfnamefont {S.}~\bibnamefont {Holleis}},
  \bibinfo {author} {\bibfnamefont {M.~W.}\ \bibnamefont {Rupich}}, \bibinfo
  {author} {\bibfnamefont {S.}~\bibnamefont {Sathyamurthy}}, \bibinfo {author}
  {\bibfnamefont {S.}~\bibnamefont {Fleshler}}, \bibinfo {author}
  {\bibfnamefont {H.~P.}\ \bibnamefont {Sheng}}, \bibinfo {author}
  {\bibfnamefont {D.~J.}\ \bibnamefont {Miller}}, \bibinfo {author}
  {\bibfnamefont {S.}~\bibnamefont {Eley}}, \bibinfo {author} {\bibfnamefont
  {L.}~\bibnamefont {Civale}}, \bibinfo {author} {\bibfnamefont
  {A.}~\bibnamefont {Kayani}}, \bibinfo {author} {\bibfnamefont {P.~M.}\
  \bibnamefont {Niraula}}, \bibinfo {author} {\bibfnamefont {U.}~\bibnamefont
  {Welp}}, \ and\ \bibinfo {author} {\bibfnamefont {W.-K.}\ \bibnamefont
  {Kwok}},\ }\bibfield  {title} {\emph {\bibinfo {title} {{Rapid doubling of
  the critical current of
  {$\mathrm{Y}\mathrm{Ba}_2\mathrm{Cu}_3\mathrm{O}_{7-\delta}$} coated
  conductors for viable high-speed industrial processing}}},\ }\href {\doibase
  10.1063/1.4935335} {\bibfield  {journal} {\bibinfo  {journal} {Appl. Phys.
  Lett.}\ }\textbf {\bibinfo {volume} {107}},\ \bibinfo {pages} {192601}
  (\bibinfo {year} {2015})}\BibitemShut {NoStop}%
\bibitem [{\citenamefont {Taen}\ \emph {et~al.}(2012)\citenamefont {Taen},
  \citenamefont {Nakajima}, \citenamefont {Tamegai},\ and\ \citenamefont
  {Kitamura}}]{Taen2012}%
  \BibitemOpen
  \bibfield  {author} {\bibinfo {author} {\bibfnamefont {T.}~\bibnamefont
  {Taen}}, \bibinfo {author} {\bibfnamefont {Y.}~\bibnamefont {Nakajima}},
  \bibinfo {author} {\bibfnamefont {T.}~\bibnamefont {Tamegai}}, \ and\
  \bibinfo {author} {\bibfnamefont {H.}~\bibnamefont {Kitamura}},\ }\bibfield
  {title} {\emph {\bibinfo {title} {Enhancement of critical current density and
  vortex activation energy in proton-irradiated {$\mathrm{Co}$}-doped
  {$\mathrm{Ba}\mathrm{Fe}_2\mathrm{As}_2$}}},\ }\href {\doibase
  10.1103/PhysRevB.86.094527} {\bibfield  {journal} {\bibinfo  {journal} {Phys.
  Rev. B}\ }\textbf {\bibinfo {volume} {86}},\ \bibinfo {pages} {094527}
  (\bibinfo {year} {2012})}\BibitemShut {NoStop}%
\bibitem [{\citenamefont {Taen}\ \emph {et~al.}(2015)\citenamefont {Taen},
  \citenamefont {Ohtake}, \citenamefont {Pyon}, \citenamefont {Tamegai},\ and\
  \citenamefont {Kitamura}}]{Taen2015}%
  \BibitemOpen
  \bibfield  {author} {\bibinfo {author} {\bibfnamefont {T.}~\bibnamefont
  {Taen}}, \bibinfo {author} {\bibfnamefont {F.}~\bibnamefont {Ohtake}},
  \bibinfo {author} {\bibfnamefont {S.}~\bibnamefont {Pyon}}, \bibinfo {author}
  {\bibfnamefont {T.}~\bibnamefont {Tamegai}}, \ and\ \bibinfo {author}
  {\bibfnamefont {H.}~\bibnamefont {Kitamura}},\ }\bibfield  {title} {\emph
  {\bibinfo {title} {Critical current density and vortex dynamics in pristine
  and proton-irradiated
  {$\mathrm{Ba}_{0.6}\mathrm{K}_{0.4}\mathrm{Fe}_2\mathrm{As}_2$}}},\ }\href
  {\doibase 10.1088/0953-2048/28/8/085003} {\bibfield  {journal} {\bibinfo
  {journal} {Supercond. Sci. Technol.}\ }\textbf {\bibinfo {volume} {28}},\
  \bibinfo {pages} {085003} (\bibinfo {year} {2015})}\BibitemShut {NoStop}%
\bibitem [{\citenamefont {Labusch}(1969)}]{Labusch1969}%
  \BibitemOpen
  \bibfield  {author} {\bibinfo {author} {\bibfnamefont {R.}~\bibnamefont
  {Labusch}},\ }\bibfield  {title} {\emph {\bibinfo {title} {Calculation of the
  critical field gradient in type-{II} superconductors}},\ }\href@noop {}
  {\bibfield  {journal} {\bibinfo  {journal} {Cryst. Lattice Defects}\ }\textbf
  {\bibinfo {volume} {1}},\ \bibinfo {pages} {1} (\bibinfo {year}
  {1969})}\BibitemShut {NoStop}%
\bibitem [{\citenamefont {Larkin}\ and\ \citenamefont
  {Ovchinnikov}(1979)}]{LarkinO:1979}%
  \BibitemOpen
  \bibfield  {author} {\bibinfo {author} {\bibfnamefont {A.~I.}\ \bibnamefont
  {Larkin}}\ and\ \bibinfo {author} {\bibfnamefont {Y.~N.}\ \bibnamefont
  {Ovchinnikov}},\ }\bibfield  {title} {\emph {\bibinfo {title} {Pinning in
  type {II} superconductors}},\ }\href {\doibase 10.1007/BF00117160} {\bibfield
   {journal} {\bibinfo  {journal} {J. Low Temp. Phys.}\ }\textbf {\bibinfo
  {volume} {34}},\ \bibinfo {pages} {409} (\bibinfo {year} {1979})}\BibitemShut
  {NoStop}%
\bibitem [{\citenamefont {Vinokur}\ \emph {et~al.}(1990)\citenamefont
  {Vinokur}, \citenamefont {Kes},\ and\ \citenamefont
  {Koshelev}}]{Vinokur1990}%
  \BibitemOpen
  \bibfield  {author} {\bibinfo {author} {\bibfnamefont {V.~M.}\ \bibnamefont
  {Vinokur}}, \bibinfo {author} {\bibfnamefont {P.~H.}\ \bibnamefont {Kes}}, \
  and\ \bibinfo {author} {\bibfnamefont {A.~E.}\ \bibnamefont {Koshelev}},\
  }\bibfield  {title} {\emph {\bibinfo {title} {Flux pinning and creep in very
  anistropic high temperature superconductors}},\ }\href {\doibase
  10.1016/0921-4534(90)90100-S} {\bibfield  {journal} {\bibinfo  {journal}
  {Phys. C}\ }\textbf {\bibinfo {volume} {168}},\ \bibinfo {pages} {29}
  (\bibinfo {year} {1990})}\BibitemShut {NoStop}%
\bibitem [{\citenamefont {Coffey}\ and\ \citenamefont
  {Clem}(1991)}]{Coffey1991}%
  \BibitemOpen
  \bibfield  {author} {\bibinfo {author} {\bibfnamefont {M.~W.}\ \bibnamefont
  {Coffey}}\ and\ \bibinfo {author} {\bibfnamefont {J.~R.}\ \bibnamefont
  {Clem}},\ }\bibfield  {title} {\emph {\bibinfo {title} {Unified theory of
  effects of vortex pinning and flux creep upon the rf surface impedance of
  type-{II} superconductors}},\ }\href {\doibase 10.1103/PhysRevLett.67.386}
  {\bibfield  {journal} {\bibinfo  {journal} {Phys. Rev. Lett.}\ }\textbf
  {\bibinfo {volume} {67}},\ \bibinfo {pages} {386} (\bibinfo {year}
  {1991})}\BibitemShut {NoStop}%
\bibitem [{\citenamefont {Ovchinnikov}\ and\ \citenamefont
  {Ivlev}(1991)}]{OvchinnikovI:1991}%
  \BibitemOpen
  \bibfield  {author} {\bibinfo {author} {\bibfnamefont {Y.~N.}\ \bibnamefont
  {Ovchinnikov}}\ and\ \bibinfo {author} {\bibfnamefont {B.~I.}\ \bibnamefont
  {Ivlev}},\ }\bibfield  {title} {\emph {\bibinfo {title} {Pinning in layered
  inhomogeneous superconductors}},\ }\href {\doibase 10.1103/PhysRevB.43.8024}
  {\bibfield  {journal} {\bibinfo  {journal} {Phys. Rev. B}\ }\textbf {\bibinfo
  {volume} {43}},\ \bibinfo {pages} {8024} (\bibinfo {year}
  {1991})}\BibitemShut {NoStop}%
\bibitem [{\citenamefont {Blatter}\ \emph {et~al.}(1994)\citenamefont
  {Blatter}, \citenamefont {Feigel'man}, \citenamefont {Geshkenbein},
  \citenamefont {Larkin},\ and\ \citenamefont {Vinokur}}]{BlatterFGLV:1994}%
  \BibitemOpen
  \bibfield  {author} {\bibinfo {author} {\bibfnamefont {G.}~\bibnamefont
  {Blatter}}, \bibinfo {author} {\bibfnamefont {M.~V.}\ \bibnamefont
  {Feigel'man}}, \bibinfo {author} {\bibfnamefont {V.~B.}\ \bibnamefont
  {Geshkenbein}}, \bibinfo {author} {\bibfnamefont {A.~I.}\ \bibnamefont
  {Larkin}}, \ and\ \bibinfo {author} {\bibfnamefont {V.~M.}\ \bibnamefont
  {Vinokur}},\ }\bibfield  {title} {\emph {\bibinfo {title} {Vortices in
  high-temperature superconductors}},\ }\href {\doibase
  10.1103/RevModPhys.66.1125} {\bibfield  {journal} {\bibinfo  {journal} {Rev.
  Mod. Phys.}\ }\textbf {\bibinfo {volume} {66}},\ \bibinfo {pages} {1125}
  (\bibinfo {year} {1994})}\BibitemShut {NoStop}%
\bibitem [{\citenamefont {Brandt}(1995)}]{Brandt:1995}%
  \BibitemOpen
  \bibfield  {author} {\bibinfo {author} {\bibfnamefont {E.~H.}\ \bibnamefont
  {Brandt}},\ }\bibfield  {title} {\emph {\bibinfo {title} {The
  flux-line-lattice in superconductors}},\ }\href {\doibase
  10.1088/0034-4885/58/11/003} {\bibfield  {journal} {\bibinfo  {journal} {Rep.
  Prog. Phys.}\ }\textbf {\bibinfo {volume} {58}},\ \bibinfo {pages} {1465}
  (\bibinfo {year} {1995})}\BibitemShut {NoStop}%
\bibitem [{\citenamefont {Blatter}\ \emph {et~al.}(2004)\citenamefont
  {Blatter}, \citenamefont {Geshkenbein},\ and\ \citenamefont
  {Koopmann}}]{BlatterGK:2004}%
  \BibitemOpen
  \bibfield  {author} {\bibinfo {author} {\bibfnamefont {G.}~\bibnamefont
  {Blatter}}, \bibinfo {author} {\bibfnamefont {V.~B.}\ \bibnamefont
  {Geshkenbein}}, \ and\ \bibinfo {author} {\bibfnamefont {J.~A.~G.}\
  \bibnamefont {Koopmann}},\ }\bibfield  {title} {\emph {\bibinfo {title} {Weak
  to strong pinning crossover}},\ }\href {\doibase
  10.1103/PhysRevLett.92.067009} {\bibfield  {journal} {\bibinfo  {journal}
  {Phys. Rev. Lett.}\ }\textbf {\bibinfo {volume} {92}},\ \bibinfo {pages}
  {067009} (\bibinfo {year} {2004})}\BibitemShut {NoStop}%
\bibitem [{\citenamefont {Gurevich}(2007)}]{GurevichSST07}%
  \BibitemOpen
  \bibfield  {author} {\bibinfo {author} {\bibfnamefont {A.}~\bibnamefont
  {Gurevich}},\ }\bibfield  {title} {\emph {\bibinfo {title} {Pinning size
  effects in critical currents of superconducting films}},\ }\href {\doibase
  10.1088/0953-2048/20/9/S03} {\bibfield  {journal} {\bibinfo  {journal}
  {Supercond. Sci. Technol.}\ }\textbf {\bibinfo {volume} {20}},\ \bibinfo
  {pages} {S128} (\bibinfo {year} {2007})}\BibitemShut {NoStop}%
\bibitem [{\citenamefont {Buchacek}\ \emph {et~al.}()\citenamefont {Buchacek},
  \citenamefont {Willa}, \citenamefont {Geshkenbein},\ and\ \citenamefont
  {Blatter}}]{Buchacek2018a}%
  \BibitemOpen
  \bibfield  {author} {\bibinfo {author} {\bibfnamefont {M.}~\bibnamefont
  {Buchacek}}, \bibinfo {author} {\bibfnamefont {R.}~\bibnamefont {Willa}},
  \bibinfo {author} {\bibfnamefont {V.~B.}\ \bibnamefont {Geshkenbein}}, \ and\
  \bibinfo {author} {\bibfnamefont {G.}~\bibnamefont {Blatter}},\ }\bibfield
  {title} {\emph {\bibinfo {title} {Strong pinning theory of thermal vortex
  creep in type-{II} superconductors}},\ }\href@noop {} {\ }\Eprint
  {http://arxiv.org/abs/1802.00652} {arXiv:1802.00652} \BibitemShut {NoStop}%
\bibitem [{\citenamefont {Brandt}(1983{\natexlab{a}})}]{BrandtJLTP83-1}%
  \BibitemOpen
  \bibfield  {author} {\bibinfo {author} {\bibfnamefont {E.}~\bibnamefont
  {Brandt}},\ }\bibfield  {title} {\emph {\bibinfo {title} {Computer simulation
  of vortex pinning in type {II} superconductors. {I}. two-dimensional
  simulation}},\ }\href {\doibase 10.1007/BF00685776} {\bibfield  {journal}
  {\bibinfo  {journal} {J. Low Temp. Phys.}\ }\textbf {\bibinfo {volume}
  {53}},\ \bibinfo {pages} {41} (\bibinfo {year}
  {1983}{\natexlab{a}})}\BibitemShut {NoStop}%
\bibitem [{\citenamefont {Brandt}(1983{\natexlab{b}})}]{BrandtJLTP83-2}%
  \BibitemOpen
  \bibfield  {author} {\bibinfo {author} {\bibfnamefont {E.}~\bibnamefont
  {Brandt}},\ }\bibfield  {title} {\emph {\bibinfo {title} {Computer simulation
  of vortex pinning in type {II} superconductors. {II}. random point pins}},\
  }\href {\doibase 10.1007/BF00685777} {\bibfield  {journal} {\bibinfo
  {journal} {J. Low Temp. Phys.}\ }\textbf {\bibinfo {volume} {53}},\ \bibinfo
  {pages} {71} (\bibinfo {year} {1983}{\natexlab{b}})}\BibitemShut {NoStop}%
\bibitem [{\citenamefont {Jensen}\ \emph {et~al.}(1988)\citenamefont {Jensen},
  \citenamefont {Brass},\ and\ \citenamefont
  {Berlinsky}}]{JensenPhysRevLett88}%
  \BibitemOpen
  \bibfield  {author} {\bibinfo {author} {\bibfnamefont {H.~J.}\ \bibnamefont
  {Jensen}}, \bibinfo {author} {\bibfnamefont {A.}~\bibnamefont {Brass}}, \
  and\ \bibinfo {author} {\bibfnamefont {A.~J.}\ \bibnamefont {Berlinsky}},\
  }\bibfield  {title} {\emph {\bibinfo {title} {Lattice deformations and
  plastic flow through bottlenecks in a two-dimensional model for flux pinning
  in type-{II} superconductors}},\ }\href {\doibase
  10.1103/PhysRevLett.60.1676} {\bibfield  {journal} {\bibinfo  {journal}
  {Phys. Rev. Lett.}\ }\textbf {\bibinfo {volume} {60}},\ \bibinfo {pages}
  {1676} (\bibinfo {year} {1988})}\BibitemShut {NoStop}%
\bibitem [{\citenamefont {Koshelev}\ and\ \citenamefont
  {Vinokur}(1994)}]{KoshelevPhysRevLett94}%
  \BibitemOpen
  \bibfield  {author} {\bibinfo {author} {\bibfnamefont {A.~E.}\ \bibnamefont
  {Koshelev}}\ and\ \bibinfo {author} {\bibfnamefont {V.~M.}\ \bibnamefont
  {Vinokur}},\ }\bibfield  {title} {\emph {\bibinfo {title} {Dynamic melting of
  the vortex lattice}},\ }\href {\doibase 10.1103/PhysRevLett.73.3580}
  {\bibfield  {journal} {\bibinfo  {journal} {Phys. Rev. Lett.}\ }\textbf
  {\bibinfo {volume} {73}},\ \bibinfo {pages} {3580} (\bibinfo {year}
  {1994})}\BibitemShut {NoStop}%
\bibitem [{\citenamefont {van Otterlo}\ \emph {et~al.}(2000)\citenamefont {van
  Otterlo}, \citenamefont {Scalettar}, \citenamefont {Zim\'anyi}, \citenamefont
  {Olsson}, \citenamefont {Petrean}, \citenamefont {Kwok},\ and\ \citenamefont
  {Vinokur}}]{OtterloPRL00}%
  \BibitemOpen
  \bibfield  {author} {\bibinfo {author} {\bibfnamefont {A.}~\bibnamefont {van
  Otterlo}}, \bibinfo {author} {\bibfnamefont {R.~T.}\ \bibnamefont
  {Scalettar}}, \bibinfo {author} {\bibfnamefont {G.~T.}\ \bibnamefont
  {Zim\'anyi}}, \bibinfo {author} {\bibfnamefont {R.}~\bibnamefont {Olsson}},
  \bibinfo {author} {\bibfnamefont {A.}~\bibnamefont {Petrean}}, \bibinfo
  {author} {\bibfnamefont {W.}~\bibnamefont {Kwok}}, \ and\ \bibinfo {author}
  {\bibfnamefont {V.}~\bibnamefont {Vinokur}},\ }\bibfield  {title} {\emph
  {\bibinfo {title} {Dynamic phases and the peak effect in dirty type {II}
  superconductors}},\ }\href {\doibase 10.1103/PhysRevLett.84.2493} {\bibfield
  {journal} {\bibinfo  {journal} {Phys. Rev. Lett.}\ }\textbf {\bibinfo
  {volume} {84}},\ \bibinfo {pages} {2493} (\bibinfo {year}
  {2000})}\BibitemShut {NoStop}%
\bibitem [{\citenamefont {Winiecki}\ and\ \citenamefont
  {Adams}(2002)}]{WinieckiA:2002}%
  \BibitemOpen
  \bibfield  {author} {\bibinfo {author} {\bibfnamefont {T.}~\bibnamefont
  {Winiecki}}\ and\ \bibinfo {author} {\bibfnamefont {C.~S.}\ \bibnamefont
  {Adams}},\ }\bibfield  {title} {\emph {\bibinfo {title} {{Time-dependent
  Ginzburg-Landau simulations of the voltage-current characteristic of type-II
  superconductors with pinning}}},\ }\href {\doibase
  10.1103/PhysRevB.65.104517} {\bibfield  {journal} {\bibinfo  {journal} {Phys.
  Rev. B}\ }\textbf {\bibinfo {volume} {65}},\ \bibinfo {pages} {104517}
  (\bibinfo {year} {2002})}\BibitemShut {NoStop}%
\bibitem [{\citenamefont {Luo}\ and\ \citenamefont {Hu}(2007)}]{LuoHu:2007}%
  \BibitemOpen
  \bibfield  {author} {\bibinfo {author} {\bibfnamefont {M.-B.}\ \bibnamefont
  {Luo}}\ and\ \bibinfo {author} {\bibfnamefont {X.}~\bibnamefont {Hu}},\
  }\bibfield  {title} {\emph {\bibinfo {title} {Depinning and creep motion in
  glass states of flux lines}},\ }\href {\doibase
  10.1103/PhysRevLett.98.267002} {\bibfield  {journal} {\bibinfo  {journal}
  {Phys. Rev. Lett.}\ }\textbf {\bibinfo {volume} {98}},\ \bibinfo {pages}
  {267002} (\bibinfo {year} {2007})}\BibitemShut {NoStop}%
\bibitem [{\citenamefont {Luo}\ and\ \citenamefont {Hu}(2010)}]{LuoHuJSNM10}%
  \BibitemOpen
  \bibfield  {author} {\bibinfo {author} {\bibfnamefont {M.-B.}\ \bibnamefont
  {Luo}}\ and\ \bibinfo {author} {\bibfnamefont {X.}~\bibnamefont {Hu}},\
  }\bibfield  {title} {\emph {\bibinfo {title} {Creep of driven flux lines in
  type-{II} superconductors}},\ }\href {\doibase 10.1007/s10948-010-0757-1}
  {\bibfield  {journal} {\bibinfo  {journal} {Journal of Superconductivity and
  Novel Magnetism}\ }\textbf {\bibinfo {volume} {23}},\ \bibinfo {pages} {1055}
  (\bibinfo {year} {2010})}\BibitemShut {NoStop}%
\bibitem [{\citenamefont {Koshelev}\ and\ \citenamefont
  {Kolton}(2011)}]{Koshelev:2011}%
  \BibitemOpen
  \bibfield  {author} {\bibinfo {author} {\bibfnamefont {A.~E.}\ \bibnamefont
  {Koshelev}}\ and\ \bibinfo {author} {\bibfnamefont {A.~B.}\ \bibnamefont
  {Kolton}},\ }\bibfield  {title} {\emph {\bibinfo {title} {Theory and
  simulations on strong pinning of vortex lines by nanoparticles}},\ }\href
  {\doibase 10.1103/PhysRevB.84.104528} {\bibfield  {journal} {\bibinfo
  {journal} {Phys. Rev. B}\ }\textbf {\bibinfo {volume} {84}},\ \bibinfo
  {pages} {104528} (\bibinfo {year} {2011})}\BibitemShut {NoStop}%
\bibitem [{\citenamefont {Dobramysl}\ \emph {et~al.}(2013)\citenamefont
  {Dobramysl}, \citenamefont {Assi}, \citenamefont {Pleimling},\ and\
  \citenamefont {T\"{a}uber}}]{DobramyslEPJ13}%
  \BibitemOpen
  \bibfield  {author} {\bibinfo {author} {\bibfnamefont {U.}~\bibnamefont
  {Dobramysl}}, \bibinfo {author} {\bibfnamefont {H.}~\bibnamefont {Assi}},
  \bibinfo {author} {\bibfnamefont {M.}~\bibnamefont {Pleimling}}, \ and\
  \bibinfo {author} {\bibfnamefont {U.}~\bibnamefont {T\"{a}uber}},\ }\bibfield
   {title} {\emph {\bibinfo {title} {Relaxation dynamics in type-{II}
  superconductors with point-like and correlated disorder}},\ }\href {\doibase
  10.1140/epjb/e2013-31101-x} {\bibfield  {journal} {\bibinfo  {journal} {Eur.
  Phys. J. B}\ }\textbf {\bibinfo {volume} {86}},\ \bibinfo {pages} {228}
  (\bibinfo {year} {2013})}\BibitemShut {NoStop}%
\bibitem [{\citenamefont {Koshelev}\ \emph {et~al.}(2016)\citenamefont
  {Koshelev}, \citenamefont {Sadovskyy}, \citenamefont {Phillips},\ and\
  \citenamefont {Glatz}}]{KoshelevPRB16}%
  \BibitemOpen
  \bibfield  {author} {\bibinfo {author} {\bibfnamefont {A.~E.}\ \bibnamefont
  {Koshelev}}, \bibinfo {author} {\bibfnamefont {I.~A.}\ \bibnamefont
  {Sadovskyy}}, \bibinfo {author} {\bibfnamefont {C.~L.}\ \bibnamefont
  {Phillips}}, \ and\ \bibinfo {author} {\bibfnamefont {A.}~\bibnamefont
  {Glatz}},\ }\bibfield  {title} {\emph {\bibinfo {title} {Optimization of
  vortex pinning by nanoparticles using simulations of the time-dependent
  {Ginzburg-Landau} model}},\ }\href {\doibase 10.1103/PhysRevB.93.060508}
  {\bibfield  {journal} {\bibinfo  {journal} {Phys. Rev. B}\ }\textbf {\bibinfo
  {volume} {93}},\ \bibinfo {pages} {060508} (\bibinfo {year}
  {2016})}\BibitemShut {NoStop}%
\bibitem [{\citenamefont {Willa}\ \emph {et~al.}(2018)\citenamefont {Willa},
  \citenamefont {Koshelev}, \citenamefont {Sadovskyy},\ and\ \citenamefont
  {Glatz}}]{Willa2018a}%
  \BibitemOpen
  \bibfield  {author} {\bibinfo {author} {\bibfnamefont {R.}~\bibnamefont
  {Willa}}, \bibinfo {author} {\bibfnamefont {A.~E.}\ \bibnamefont {Koshelev}},
  \bibinfo {author} {\bibfnamefont {I.~A.}\ \bibnamefont {Sadovskyy}}, \ and\
  \bibinfo {author} {\bibfnamefont {A.}~\bibnamefont {Glatz}},\ }\bibfield
  {title} {\emph {\bibinfo {title} {Strong-pinning regimes by spherical
  inclusions in anisotropic type-{II} superconductors}},\ }\href {\doibase
  10.1088/1361-6668/aa939e} {\bibfield  {journal} {\bibinfo  {journal}
  {Supercond. Sci. Technol.}\ }\textbf {\bibinfo {volume} {31}},\ \bibinfo
  {pages} {014001} (\bibinfo {year} {2018})}\BibitemShut {NoStop}%
\bibitem [{\citenamefont {Sadovskyy}\ \emph {et~al.}(2015)\citenamefont
  {Sadovskyy}, \citenamefont {Koshelev}, \citenamefont {Phillips},
  \citenamefont {Karpeyev},\ and\ \citenamefont {Glatz}}]{SadovskyyJComp2015}%
  \BibitemOpen
  \bibfield  {author} {\bibinfo {author} {\bibfnamefont {I.}~\bibnamefont
  {Sadovskyy}}, \bibinfo {author} {\bibfnamefont {A.}~\bibnamefont {Koshelev}},
  \bibinfo {author} {\bibfnamefont {C.}~\bibnamefont {Phillips}}, \bibinfo
  {author} {\bibfnamefont {D.}~\bibnamefont {Karpeyev}}, \ and\ \bibinfo
  {author} {\bibfnamefont {A.}~\bibnamefont {Glatz}},\ }\bibfield  {title}
  {\emph {\bibinfo {title} {Stable large-scale solver for {Ginzburg-Landau}
  equations for superconductors}},\ }\href {\doibase 10.1016/j.jcp.2015.04.002}
  {\bibfield  {journal} {\bibinfo  {journal} {J. Comput. Phys.}\ }\textbf
  {\bibinfo {volume} {294}},\ \bibinfo {pages} {639} (\bibinfo {year}
  {2015})}\BibitemShut {NoStop}%
\bibitem [{\citenamefont {Sadovskyy}\ \emph
  {et~al.}(2016{\natexlab{a}})\citenamefont {Sadovskyy}, \citenamefont
  {Koshelev}, \citenamefont {Glatz}, \citenamefont {Ortalan}, \citenamefont
  {Rupich},\ and\ \citenamefont {Leroux}}]{SadovskyyPRAppl2016}%
  \BibitemOpen
  \bibfield  {author} {\bibinfo {author} {\bibfnamefont {I.~A.}\ \bibnamefont
  {Sadovskyy}}, \bibinfo {author} {\bibfnamefont {A.~E.}\ \bibnamefont
  {Koshelev}}, \bibinfo {author} {\bibfnamefont {A.}~\bibnamefont {Glatz}},
  \bibinfo {author} {\bibfnamefont {V.}~\bibnamefont {Ortalan}}, \bibinfo
  {author} {\bibfnamefont {M.~W.}\ \bibnamefont {Rupich}}, \ and\ \bibinfo
  {author} {\bibfnamefont {M.}~\bibnamefont {Leroux}},\ }\bibfield  {title}
  {\emph {\bibinfo {title} {Simulation of the vortex dynamics in a real pinning
  landscape of
  ${\mathrm{yba}}_{2}{\mathrm{cu}}_{3}{\mathrm{o}}_{7\ensuremath{-}\ensuremath{\delta}}$
  coated conductors}},\ }\href {\doibase 10.1103/PhysRevApplied.5.014011}
  {\bibfield  {journal} {\bibinfo  {journal} {Phys. Rev. Applied}\ }\textbf
  {\bibinfo {volume} {5}},\ \bibinfo {pages} {014011} (\bibinfo {year}
  {2016}{\natexlab{a}})}\BibitemShut {NoStop}%
\bibitem [{\citenamefont {Sadovskyy}\ \emph
  {et~al.}(2016{\natexlab{b}})\citenamefont {Sadovskyy}, \citenamefont {Jia},
  \citenamefont {Leroux}, \citenamefont {Kwon}, \citenamefont {Hu},
  \citenamefont {Fang}, \citenamefont {Chaparro}, \citenamefont {Zhu},
  \citenamefont {Welp}, \citenamefont {Zuo}, \citenamefont {Zhang},
  \citenamefont {Nakasaki}, \citenamefont {Selvamanickam}, \citenamefont
  {Crabtree}, \citenamefont {Koshelev}, \citenamefont {Glatz},\ and\
  \citenamefont {Kwok}}]{SadovskyyAdvMat2016}%
  \BibitemOpen
  \bibfield  {author} {\bibinfo {author} {\bibfnamefont {I.~A.}\ \bibnamefont
  {Sadovskyy}}, \bibinfo {author} {\bibfnamefont {Y.}~\bibnamefont {Jia}},
  \bibinfo {author} {\bibfnamefont {M.}~\bibnamefont {Leroux}}, \bibinfo
  {author} {\bibfnamefont {J.}~\bibnamefont {Kwon}}, \bibinfo {author}
  {\bibfnamefont {H.}~\bibnamefont {Hu}}, \bibinfo {author} {\bibfnamefont
  {L.}~\bibnamefont {Fang}}, \bibinfo {author} {\bibfnamefont {C.}~\bibnamefont
  {Chaparro}}, \bibinfo {author} {\bibfnamefont {S.}~\bibnamefont {Zhu}},
  \bibinfo {author} {\bibfnamefont {U.}~\bibnamefont {Welp}}, \bibinfo {author}
  {\bibfnamefont {J.-M.}\ \bibnamefont {Zuo}}, \bibinfo {author} {\bibfnamefont
  {Y.}~\bibnamefont {Zhang}}, \bibinfo {author} {\bibfnamefont
  {R.}~\bibnamefont {Nakasaki}}, \bibinfo {author} {\bibfnamefont
  {V.}~\bibnamefont {Selvamanickam}}, \bibinfo {author} {\bibfnamefont {G.~W.}\
  \bibnamefont {Crabtree}}, \bibinfo {author} {\bibfnamefont {A.~E.}\
  \bibnamefont {Koshelev}}, \bibinfo {author} {\bibfnamefont {A.}~\bibnamefont
  {Glatz}}, \ and\ \bibinfo {author} {\bibfnamefont {W.-K.}\ \bibnamefont
  {Kwok}},\ }\bibfield  {title} {\emph {\bibinfo {title} {Toward
  superconducting critical current by design}},\ }\href {\doibase
  10.1002/adma.201600602} {\bibfield  {journal} {\bibinfo  {journal} {Advanced
  Materials}\ }\textbf {\bibinfo {volume} {28}},\ \bibinfo {pages} {4593}
  (\bibinfo {year} {2016}{\natexlab{b}})}\BibitemShut {NoStop}%
\bibitem [{\citenamefont {Kimmel}\ \emph {et~al.}(2017)\citenamefont {Kimmel},
  \citenamefont {Sadovskyy},\ and\ \citenamefont {Glatz}}]{KimmelPRE2017}%
  \BibitemOpen
  \bibfield  {author} {\bibinfo {author} {\bibfnamefont {G.}~\bibnamefont
  {Kimmel}}, \bibinfo {author} {\bibfnamefont {I.~A.}\ \bibnamefont
  {Sadovskyy}}, \ and\ \bibinfo {author} {\bibfnamefont {A.}~\bibnamefont
  {Glatz}},\ }\bibfield  {title} {\emph {\bibinfo {title} {In silico
  optimization of critical currents in superconductors}},\ }\href {\doibase
  10.1103/PhysRevE.96.013318} {\bibfield  {journal} {\bibinfo  {journal} {Phys.
  Rev. E}\ }\textbf {\bibinfo {volume} {96}},\ \bibinfo {pages} {013318}
  (\bibinfo {year} {2017})}\BibitemShut {NoStop}%
\bibitem [{Sup()}]{SupplMat-arxiv}%
  \BibitemOpen
  \href@noop {} {}\bibinfo {note} {{Supplementary Material includes animations
  of a single vortex or vortex lattice detaching from a single defect:\\
  \uhref{https://youtu.be/dRf0XTxe14M}{Vortex depinning from large defect [$a =
  4\xi$]}\\ \uhref{https://youtu.be/W_CvXXTWrMA}{Vortex depinning from very
  large defect [$a = 16\xi$]}\\ \uhref{https://youtu.be/BqOB8YRs4Lo}{V. lattice
  depinning from defect [$a=4\xi$, $B = 0.071 H_{c2}$]}\\
  \uhref{https://youtu.be/npn-LTNn5U8}{V. lattice depinning from defect
  [$a=4\xi$, $B = 0.19 H_{c2}$]}\\ \uhref{https://youtu.be/3n7o5JvG0nA}{V.
  lattice depinning from defect [$a=4\xi$, $B = 0.295 H_{c2}$]}\\
  \uhref{https://youtu.be/3ITEVTkJ1pA}{V. lattice depinning from defect
  [$a=6\xi$, $B = 0.221 H_{c2}$]}\\ \uhref{https://youtu.be/moUb4VO7GZA}{V.
  lattice depinning from defect [$a=6\xi$, $B = 0.295 H_{c2}$]}\\ More
  animations at
  \uhref{https://www.youtube.com/channel/UCjdQ4Ruhxma5pkxGrFxw3CA}{OSCon-SciDAC
  YouTube channel}.}}\BibitemShut {Stop}%
\bibitem [{\citenamefont {Willa}\ \emph
  {et~al.}(2015{\natexlab{a}})\citenamefont {Willa}, \citenamefont
  {Geshkenbein}, \citenamefont {Prozorov},\ and\ \citenamefont
  {Blatter}}]{Willa2015a}%
  \BibitemOpen
  \bibfield  {author} {\bibinfo {author} {\bibfnamefont {R.}~\bibnamefont
  {Willa}}, \bibinfo {author} {\bibfnamefont {V.~B.}\ \bibnamefont
  {Geshkenbein}}, \bibinfo {author} {\bibfnamefont {R.}~\bibnamefont
  {Prozorov}}, \ and\ \bibinfo {author} {\bibfnamefont {G.}~\bibnamefont
  {Blatter}},\ }\bibfield  {title} {\emph {\bibinfo {title} {{Campbell}
  response in type-{II} superconductors under strong pinning conditions}},\
  }\href {\doibase 10.1103/PhysRevLett.115.207001} {\bibfield  {journal}
  {\bibinfo  {journal} {Phys. Rev. Lett.}\ }\textbf {\bibinfo {volume} {115}},\
  \bibinfo {pages} {207001} (\bibinfo {year} {2015}{\natexlab{a}})}\BibitemShut
  {NoStop}%
\bibitem [{\citenamefont {Willa}\ \emph {et~al.}(2016)\citenamefont {Willa},
  \citenamefont {Geshkenbein},\ and\ \citenamefont {Blatter}}]{Willa2016}%
  \BibitemOpen
  \bibfield  {author} {\bibinfo {author} {\bibfnamefont {R.}~\bibnamefont
  {Willa}}, \bibinfo {author} {\bibfnamefont {V.~B.}\ \bibnamefont
  {Geshkenbein}}, \ and\ \bibinfo {author} {\bibfnamefont {G.}~\bibnamefont
  {Blatter}},\ }\bibfield  {title} {\emph {\bibinfo {title} {Probing the
  pinning landscape in type-{II} superconductors via {Campbell} penetration
  depth}},\ }\href {\doibase 10.1103/PhysRevB.93.064515} {\bibfield  {journal}
  {\bibinfo  {journal} {Phys. Rev. B}\ }\textbf {\bibinfo {volume} {93}},\
  \bibinfo {pages} {064515} (\bibinfo {year} {2016})}\BibitemShut {NoStop}%
\bibitem [{\citenamefont {Brandt}(1977{\natexlab{a}})}]{Brandt1977a}%
  \BibitemOpen
  \bibfield  {author} {\bibinfo {author} {\bibfnamefont {E.~H.}\ \bibnamefont
  {Brandt}},\ }\bibfield  {title} {\emph {\bibinfo {title} {{Elastic energy of
  the vortex state in type-II superconductors. I. High inductions}}},\
  }\href@noop {} {\bibfield  {journal} {\bibinfo  {journal} {J. Low Temp.
  Phys.}\ }\textbf {\bibinfo {volume} {26}},\ \bibinfo {pages} {709} (\bibinfo
  {year} {1977}{\natexlab{a}})}\BibitemShut {NoStop}%
\bibitem [{\citenamefont {Brandt}(1977{\natexlab{b}})}]{Brandt1977b}%
  \BibitemOpen
  \bibfield  {author} {\bibinfo {author} {\bibfnamefont {E.~H.}\ \bibnamefont
  {Brandt}},\ }\bibfield  {title} {\emph {\bibinfo {title} {{Elastic energy of
  the vortex state in type-II superconductors. II. Low inductions}}},\
  }\href@noop {} {\bibfield  {journal} {\bibinfo  {journal} {J. Low Temp.
  Phys.}\ }\textbf {\bibinfo {volume} {26}},\ \bibinfo {pages} {735} (\bibinfo
  {year} {1977}{\natexlab{b}})}\BibitemShut {NoStop}%
\bibitem [{\citenamefont {Brandt}(1986)}]{Brandt1986}%
  \BibitemOpen
  \bibfield  {author} {\bibinfo {author} {\bibfnamefont {E.~H.}\ \bibnamefont
  {Brandt}},\ }\bibfield  {title} {\emph {\bibinfo {title} {Elastic and plastic
  properties of the flux-line lattice in type-{II} superconductors}},\ }\href
  {\doibase 10.1103/PhysRevB.34.6514} {\bibfield  {journal} {\bibinfo
  {journal} {Phys. Rev. B}\ }\textbf {\bibinfo {volume} {34}},\ \bibinfo
  {pages} {6514} (\bibinfo {year} {1986})}\BibitemShut {NoStop}%
\bibitem [{\citenamefont {Willa}\ \emph
  {et~al.}(2015{\natexlab{b}})\citenamefont {Willa}, \citenamefont
  {Geshkenbein},\ and\ \citenamefont {Blatter}}]{Willa2015b}%
  \BibitemOpen
  \bibfield  {author} {\bibinfo {author} {\bibfnamefont {R.}~\bibnamefont
  {Willa}}, \bibinfo {author} {\bibfnamefont {V.~B.}\ \bibnamefont
  {Geshkenbein}}, \ and\ \bibinfo {author} {\bibfnamefont {G.}~\bibnamefont
  {Blatter}},\ }\bibfield  {title} {\emph {\bibinfo {title} {{Campbell}
  penetration in the critical state of type-{II} superconductors}},\ }\href
  {\doibase 10.1103/PhysRevB.92.134501} {\bibfield  {journal} {\bibinfo
  {journal} {Phys. Rev. B}\ }\textbf {\bibinfo {volume} {92}},\ \bibinfo
  {pages} {134501} (\bibinfo {year} {2015}{\natexlab{b}})}\BibitemShut
  {NoStop}%
\end{thebibliography}%

\end{document}